\documentclass[aoas,nameyear,MSNbibl,secthm,seceqn,rotating,dvips]{arximspdf}
\usepackage{tikz}
\usepackage{stfloats}
\usepackage{dcolumn}
\usepackage{graphicx}

\doi{10.1214/10-AOAS447}
\volume{5}
\issue{2B}
\pubyear{2011}
\firstpage{1553}
\lastpage{1585}

\makeatletter

\fnbelowfloat

\newcommand{\cab}{C_{ab}}
\newcommand{\argmin}{\operatorname{argmin}}
\newcommand{\xB}{x^B}
\newcommand{\binom}[2]{{#1\choose #2}}

\newcolumntype{d}[1]{D{.}{.}{#1}}

  \let\sv@tabnotetext\tabnotetext
  \let\sv@tabnotemark@fmt\tabnotemark@fmt
   \long\def\legend#1{{\let\tabnote@indent\leavevmode\sv@tabnotetext[]{}{#1}}}

\def\@bmisc[#1]{%
  \get@battribute{unstr}%
  \common@pub@types%
  \let\bauthor\bbl@bauthor%
  \let\bhowpublished\@firstofone%
  \def\borganization##1{{\bauthor@style ##1}}%
}

\makeatother

\begin{document}
\begin{frontmatter}

\title{A hierarchical Bayesian approach to record linkage and
population size problems\thanksref{T1}}
\runtitle{Bayesian matching and population size}
\thankstext{T1}{Supported by MIUR Grant 2008
CEFF37;002, Bayesian Methods for finite populations, Italy.}

\begin{aug}
\author[A]{\fnms{Andrea} \snm{Tancredi}\ead[label=e1]{andrea.tancredi@uniroma1.it}}
and
\author[A]{\fnms{Brunero} \snm{Liseo}\corref{}\ead[label=e2]{brunero.liseo@uniroma1.it}}
\runauthor{A. Tancredi and B. Liseo}
\affiliation{Sapienza Universit\`{a} di Roma }

\address[A]{Department of Methods and Models\\
\quad for Economics, Territory and Finance\\
Sapienza Universit\`{a} di Roma\\
Viale del Castro Laurenziano 9, Roma 00161\\
Italy\\
\printead{e1}\\
\phantom{E-mail:\ }\printead*{e2}} 
\end{aug}

\received{\smonth{5} \syear{2009}}
\revised{\smonth{11} \syear{2010}}

%
\begin{abstract}
We propose and illustrate a hierarchical Bayesian approach for
matching statistical records observed on different occasions. We show
how this model can be profitably adopted both in record linkage
problems and in capture--recapture setups, where the size of a finite
population is the real object of interest. There are at least two
important differences between the proposed model-based approach and the
current practice in record linkage. First, the statistical model is
built up on the actually observed categorical variables and no
reduction (to 0--1 comparisons) of the available information takes
place. Second, the hierarchical
structure of the model allows a two-way propagation of the uncertainty
between the parameter estimation step and the matching procedure so
that no
plug-in estimates are used and the correct uncertainty is accounted for
both in estimating the population size and in performing the record linkage.
We illustrate and motivate our proposal
through a real data example and simulations.
\end{abstract}

%
\begin{keyword}
\kwd{Capture--recapture methods}
\kwd{conditional independence}
\kwd{Gibbs sampling}
\kwd{Metropolis--Hastings}
\kwd{record linkage}.
\end{keyword}

\end{frontmatter}

\setcounter{footnote}{1}

\section{Introduction}
\label{Introduction}
The current explosion in the availability of data from multiple sources,
and the relative ease of information storage have led to a~great
popularity of statistical methods which aim at merging and/or matching
statistical information available from different sources. Among these
methods, record linkage refers to the problem of identifying
statistical units which may be present in more than one data set.
\citet{mvf09} review the relevance of record linkage procedures in
official statistics and highlight the significant intertwins with
missing data and multiple systems estimation literature.

The gist of this paper is the proposal of a hierarchical Bayesian
framework which can be profitably adopted both in record linkage
problems and in capture--recapture scenarios, where the size of a
finite population is the main object of interest and the number of
``re-captured'' individuals is unknown. Most of the current
approaches to population size estimation with matching uncertainty
consider the matching and the size estimation as two logically well
separated steps. Remarkable recent exceptions are \citet{link} and
\citet{wright09} where genotype misidentification is embedded into
multiple mark-recapture models for estimating animal abundance using
DNA samples. More generally, in this paper, we propose a unified
framework where matching uncertainty is naturally accounted for in
estimating population size by using samples of multivariate
categorical variables.\looseness=-1

To motivate our approach, consider the following example, which is a
part of a real application. Suppose we have two data sets which we
call $A$ and $B$ with sizes, respectively, 34 and 45. Data set $A$
comprises all the foreign residents observed in a small census block
during the 2001 Italian census population survey (CPS). Data set $B$
comprises all the resident foreigners observed in the same census
block during the post enumeration survey (PES)\footnote{PES is usually
performed some time after CPS, to evaluate the effective coverage of
CPS.}. Both data sets report, among others, the following variables:
(1) first two consonants of the family name, (2) gender and (3) education
level. Assume that the three variables represent the only available
information to perform the match; assume also that the goal is the
estimation of~$N$, the total number of foreign residents in the census
block. The usual approach to this problem would be to search for the
pairs of units, belonging to different files, which agree perfectly on
each observed variable. In our example there are 25 pairs which show
a complete agreement. If we assume that we actually observed 25
recaptures, such information can be used easily in a capture--recapture
model to make inference on $N$. However, two complications may
arise. First, it might be possible that two different units genuinely
agree on each variable. Second, because of measurement error, observed
records for the same unit might be different in the two sampling
occasions. They could also agree as before, even if they refer to
different units with different true values. We will discuss this
example below in more detail. For the moment, Table \ref{table1} summarizes, for
the different choices of the declared number $T$ of recaptures, the
posterior distribution of $N$ assuming a noninformative prior
$p(N)\propto1/N^2$ and a hypergeometric likelihood function $p(T|N)
\propto\binom{n^A}{T} \binom{N-n^A}{n^B-T}/\binom{N}{n^B}$ with
$n^A=34$ and $n^B=45$. One can see that slightly different choices of
$T$ may produce dramatically different posterior
distributions.
\begin{table}
\tablewidth=310pt
\caption{Posterior quantiles for $N$ with the distribution
$p(N|T)\propto\binom{n^A}{T} \binom{N-n^A}{n^B-T}/\binom{N}{n^B}
\times(1/N^2) $ with $n^A=34$, $n^B=45$ and different choices of
$T$}\label{table1}
\begin{tabular*}{\tablewidth}{@{\extracolsep{\fill}}lccccccc@{}}
\hline
&\multicolumn{7}{c@{}}{$\bolds{T}$}\\[-5pt]
&\multicolumn{7}{c@{}}{\hrulefill}\\
& \textbf{24} & \textbf{25} & \textbf{26} & \textbf{27} & \textbf{28} & \textbf{29} & \textbf{30} \\
\hline
\hphantom{0}2.5\% & 57 & 56 & 54 & 53 & 51 & 50 & 49 \\
50\% & 64 & 62 & 59 & 57 & 55 & 53 & 51 \\
97.5\% & 78 & 74 & 70 & 66 & 63 & 60 & 57 \\
\hline
\end{tabular*}
\end{table}

Accounting for matching uncertainty has relevance well beyond size
estimation problems and relates to the more general problem of
inference with integrated data; see \citet{judson}. In this context, an
important exemplification is provided by \citet{larsenlahiri} who take
into account linkage uncertainty in the framework of the linear
regression model when the response variable and the covariates are
recorded on two different occasions. However, our approach can also
be applied when there is not yet a scheduled statistical analysis to
be performed on the linked data, but the linkage procedure is just the
initial step to obtain a larger and integrated reference data set.

Statistical methods for finding entries related to the same entity in
two or more files are employed in many different disciplines, such as
medicine, business administration and official statistics
[see, e.g., \citet{herzogetal}]. In these contexts it may happen that a
unique data set with all the necessary information for a particular
statistical analysis is not available. Furthermore, time and cost
constraints may make it unfeasible to create such a data set
anew. Integration at the unit level of different data sets (sample
surveys and/or administrative data sets) may be an answer to this kind
of problem. A considerable difficulty in this context is represented
by the lack of a unique identifier in the different data sets for each
unit of interest. In fact, when a set of observed variables (key
variables, henceforth) may be used as an identifier for connecting
records that refer to the same unit, particular attention should be
paid to errors, as we have seen in the introductory example, and
missing values.

To handle the record linkage process, many different methodologies
have been introduced. Some methods are na\"{\i}ve, or heuristic, that
is,\
are based only on common sense [e.g., the ``iterative method''
described in \citet{arm93}]. In a fundamental paper, \citet{fesu69}
put these kinds of problems into a firm, model based, statistical
framework. Further advances were described in a number of papers in the
1980s and 1990s: among others, \citet{jaro89}, \citet{wink93} and
\citet{br95}. \citet{laru01} introduce the representation of the
record linkage problem in terms of the mixture model [see also
\citet{lar99}]: this idea has been exploited in many other papers; see,
for example, \citet{Fortini}, \citet{mg04} and \citet{lar05} who
tackle the problem from a Bayesian perspective. All of these papers
assume that each single comparison between records in two different
files provides new information, independently of the other
comparisons. This assumption, as noted by \citet{Kelley}, is
fundamentally unsound, as illustrated in Section \ref{history}. Also,
in this respect, \citet{win00} states that \textit{``\dots because the
underlying true probabilities have not been accurately estimated,
estimated error rates (of the record linkage procedure) are not
accurate.''}

An important feature of our paper is that we propose a Bayesian model
which is based on the actual observed data rather than comparisons. In
a~similar spirit, \citet{Fortini2} discussed these ideas in the simple
setting of a single continuous variable. Here we will assume that our
key variables will be discrete, as almost always happens in
practice.

Record linkage is not the only statistical problem where matching
issues are concerned. In a bioinformatics context \citet{greenmardia}
introduce a matching matrix (very similar to our matrix $C$, see
later) into some problems of shape analysis, where configurations of
points in space need to be matched and the points are not completely
labeled.

\citet{degrootgoel} consider the situation where a random sample of
size~$n$, say, $(X_i, Z_i)$, $i = 1,\dots, n$, is drawn from a bivariate
normal distribution; however, before the sample values are recorded,
each observation $(x_i; z_i)$ gets broken into two separate
components. As a consequence, the available information is represented
by the vectors $x=(x_1,\dots, x_n)$ and $y=(y_1, \dots,
y_n)$, where $y$ is an unknown permutation of the values $(z_1, \dots,
z_n)$.

Another matching example is discussed in \citet{lindley77}, in a
forensic framework. Here the matching problem arises when some
material is found at the scene of a crime and similar material is
found on a suspect; in both cases material collection is subject to
measurement error. Lindley describes a~Bayesian method to establish
whether the two materials come from the same source or not. When
rephrasing Lindley's approach from a record linkage perspective, we
note that
that paper was the first attempt to introduce, into a Bayesian linking
model, the natural idea that two units with the same surname are more
likely to be a match if the surname is \textit{Bodolomonogoto} than if
the surname is \textit{Smith}. Similar suggestions can be found in the
seminal papers by \citet{newcombe59} and \citet{fesu69}.

The paper is structured as follows. In Section \ref{history} we
present the standard approach to record linkage. Our Bayesian approach
is discussed in Section~\ref{model}. Markov chain Monte Carlo (MCMC)
methods are needed for estimating the parameters of the model. In
Section \ref{MCMC} we describe a suitable algorithm for simulating the
posterior distribution. We also discuss a loss function approach to
the matching estimation. In Section \ref{appl} the performance of the
methodology is evaluated through a small illustrative application. A
more realistic example is shown in Section \ref{blocks}. A simulation
study is conducted in Section \ref{sim}. Finally, in Section \ref{disc}
we give a brief discussion of possible future extensions and
improvements of the method.

\section{Classic approach to record linkage}
\label{history}

Suppose we are given two record configurations $x^A$ and $x^B$ of
different sizes $n^A$ and $n^B$ with
\[
x^A=(x^A_1,\ldots,x^A_a,\ldots,x^A_{n^A})^{\prime} \quad \mbox{and}\quad
x^B=(x^B_1,\ldots,x^B_b,\ldots,x^B_{n^B})^{\prime}.
\]
Here $x^A_{a}=(x_a^{A_1},\ldots,x_a^{A_h})$ and
$x^B_{b}=(x_b^{B_1},\ldots,x_b^{B_h})$ are the observed values of a
categorical random vector $x=(x^1,\ldots,x^h)$ whose support is the
set
\[
V= \{v_{j_1 j_2,\ldots,j_h}=(v_{j_1}^1,v_{j_2}^2,\ldots,v_{j_h}^h),\
j_1=1,\ldots,k_1;\ldots; j_h=1,\ldots, k_h \}.
\]
In the following,
the two data configurations will be called, respectively, sample~$A$ and
sample $B$, the components of the random vectors $x$ 
(whenever it is possible we will avoid subscript and superscript
indices to simplify the notation) are the key variables and the
elements of the set $V$ arranged in lexicographic order will be
indicated with $v_j$ for $j=1,\ldots,k=k_1\cdot k_2\cdots k_h$.

Let $A \times B$ be the set of all possible pairs of units belonging
to different samples. Set $A \times B = M \cup U$, where $M = \{
(a,b)\in A \times B\dvtx a\equiv b \}$ (here $ a\equiv b$ means
that unit $a$ of sample $A$ and unit $b$ of sample $B$ are the same
population unit) and ${U}= \{ (a,b)\in A \times B \dvtx a\not\equiv
b \}.$ Probabilistic record linkage, as implemented, for example,
in \citet{jaro89}, is performed by modeling the comparison vectors
$y_{ab}= (y1_{ab}, \ldots, y^h_{ab})$ where
\[
y^i_{ab}= \cases{
1, &\quad $x^{A_i}_{a}=x^{B_i}_{b}$, \cr
0,&\quad $x^{A_i}_{a}\not=
x^{B_i}_{b}$,}\qquad  i=1,\ldots,h.
\]
Vectors $y_{ab}$, $a=1,\ldots,n^A$, $b=1,\ldots, n^B$, are assumed
independent conditionally on $M$ and $U$. The probability distribution
of $y_{ab}$ depends on the match or nonmatch status of the single
pair $(a,b)$; in particular, it is assumed that $ p(y_{ab}| (a,b) \in
M )= \prod_{i=1}^h m_i^{y_{ab}^i} (1-m_i)^{1-y_{ab}^i} $ and
$p(y_{ab}| (a,b) \in U )\break=\prod_{i=1}^h u_i^{y_{ab}^i}
(1-u_i)^{1-y_{ab}^i}$ (here and later, we will abuse notation by
letting the arguments define the functions) with $m=(m_1,\ldots,m_h)$
and $u=(u_1,\ldots,u_h)$ as unknown probabilities vectors. In addition,
the elements of the sets $M$ and $U$ are modeled assuming that each
pair in $A\times B$ is a match with probability $w$, independently of
all the other pairs. This way the comparison vectors $y_{ab}$ are
independent and identically distributed as a mixture of two
multivariate Bernoulli distributions:
%
\begin{eqnarray}
\label{mixbern}
p(y_{ab}|m,u,w)&=&w \prod_{i=1}^h m_i^{y_{ab}^i}
(1-m_i)^{1-y_{ab}^i}\nonumber\\[-8pt]\\[-8pt]
&&{}+(1-w) \prod_{i=1}^h u_i^{y_{ab}^i}
(1-u_i)^{1-y_{ab}^i}.\nonumber
\end{eqnarray}

Models similar to (\ref{mixbern}) are often used also in
biostatistics, under the name of a latent class model, to assess
diagnostic test accuracy in the absence of a~gold standard and only
multiple imperfect tests are available [\citet{pepe2003}]. Likelihood
maximization of the parameters in model (\ref{mixbern}) is performed
via the EM algorithm and analytical expressions for the estimators are
provided by \citet{fesu69} and \citet{pepejanes} in the case $h=3$.

Several extensions of this basic setup have been proposed; see, for
example, \citet{laru01}. In order to decide whether to declare a link a
single pair, one can consider the likelihood ratio
%
\begin{equation}
\label{FSratio} \lambda= \frac{P(y_{ab}| (a,b)\in M)}{P(y_{ab} |
(a,b) \in U)}= \frac{\prod_{i=1}^h m_i^{y_{ab}^i}
(1-m_i)^{1-y_{ab}^i}} {\prod_{i=1}^h u_i^{y_{ab}^i}
(1-u_i)^{1-y_{ab}^i}}
\end{equation}
or the posterior probability
%
\begin{eqnarray}
\label{jaropost}
&&p\bigl( (a,b) \in M |y_{ab}\bigr)\nonumber\\[-8pt]\\[-8pt]
&&\qquad =\frac{w \prod_{i=1}^h
m_i^{y_{ab}^i} (1-m_i)^{1-y_{ab}^i} }{w \prod_{i=1}^h m_i^{y_{ab}^i}
(1-m_i)^{1-y_{ab}^i} +(1-w) \prod_{i=1}^h u_i^{y_{ab}^i}
(1-u_i)^{1-y_{ab}^i}}.\nonumber
\end{eqnarray}
Pairs with high values of $\lambda$ or $p( (a,b) \in M
|y_{ab})$ are then declared matches. This approach is formalized in
the classical approach of \citet{fesu69}.

In our opinion the above approach can be criticized on several
grounds:
\begin{enumerate}
\item\textit{Decision rules for classifying records as matches.} In
general, all the pairs with a likelihood ratio $\lambda$, or a
posterior probability, above a fixed threshold are declared
matches. However, the choice of the threshold can be problematic, as
illustrated, for example, in \citet{br95}. More details about this
point will be given in Section \ref{MCMC}.

\item\textit{Avoiding multiple matches.} Current approaches to record
linkage assume that there are no duplications in the same file and
inference procedures should account for that. However, in classical
procedures, it might happen that a single record in $A$ is linked to
more than one record in $B$; consequently, some extra assumptions are
necessary. \citet{jaro89} proposes a~linear programming approach after
a preliminary match estimation step. An alternative approach
[\citet{Fortini}], which will be pursued here, incorporates the
constraints into the sampling model.

\item\textit{Incorporating sampling information.} If we assume that the
two files are random samples without replacement from a population of
unknown size $N$, an obvious prior assumption is $p((a,b) \in M)=1/N$,
with $N>\max{\{n^A,n^B\}}$. In addition, if we know that two units
assume the same value $v_j$, the matching probability becomes $p((a,b)
\in M )=1/F_j$, where~$F_j$ is the (unknown) total number of units
with record $v_j$ in the population. In record linkage procedures, in
general, sources of knowledge of this type are not included in the
model, with an obvious loss of information. This may be particularly
important for applications of record linkage in disclosure literature.

\item\textit{Comparison vectors are not independent.} In this respect
\citet{Kelley} states: \textit{``\ldots\ The
decision procedure $\dots$ was developed under the hypothesis that the
comparison vectors between separate record pairs are
independent. However, since the record pairs that are considered for
possible matches are elements of the cross product of the two files we
are attempting to match, the comparison vectors are in fact dependent
\ldots.''} As a matter of fact, the random variables $y_{ab}$ are
deterministically dependent. To see that, consider the case of one key
variable $X_1$. Suppose that $x_1^{A_1}=x_{1}^{B_1}$ and $
x_{1}^{A_1}=x_{2}^{B_1}$. If, in addition, $x_{2}^{A_1}=x_{1}^{B_1}$,
it must necessarily be true that $x_{2}^{A_1}=x_{2}^{B_1}$, that is, in
terms of comparisons, $p(y_{22}=1|y_{11}=1, y_{12}=1,
y_{21}=1)=1$. Moreover, the problem of dependency among the $y_{ab}$'s
cannot be circumvented by eliminating redundant comparisons in the
likelihood function, because the order in which pairs are considered
would matter!

\item\textit{The components $y_{ab}^i$ of the comparison vector may not
be independent conditionally on $M$ and $U$.} The conditional
independence assumption among the key variables often fails in
practice: disagreement on different key variables for a true match
might be caused by a unique reason which introduces correlation among
the $y_{ab}^i$'s. In the absence of conditional independence, the
resulting estimates of $w$, $m$ and $u$ lose their meaning and a more
sophisticated conditional dependence structure must be specified.
Similar arguments have been applied to criticize the use of model
(\ref{mixbern}) for the analysis of diagnostic test performance
without a gold standard and, in this context, several solutions have
been proposed and discussed [\citet{albertdood};
\citet{pepejanes}]. \citet{laru01} have introduced interactions among
key variables; see also \citet{winkler95} and references therein.

\end{enumerate}

\section{The new model}
\label{model}

We assume that the records in $x^A$ and $x^B$ are measurements subject
to recording error of a multivariate categorical variable
$\mu=(\mu1,\ldots,\mu^h)$ whose support is, on both occasions, the set
$V$. Specifically, let
\[
\mu^A=(\mu^A_1,\ldots,\mu^A_a,\ldots,\mu^A_{n^A})^{\prime}
\quad \mbox{and}\quad
\mu^B=(\mu^B_1,\ldots,\mu^B_b,\ldots,\mu^B_{n^B})^{\prime}
\]
be two
independent random samples from the multivariate categorical variable
$\mu$ drawn on different occasions from the same finite population.
Let $\mu_a^A=(\mu_{a}^{A_1},\ldots,\mu_{a}^{A_h}), a=1,\ldots,n^A$ and
$\mu_b^B=(\mu_{b}^{B_1},\ldots,\mu_{b}^{B_h}), b=1,\ldots,n^B$ be the
unobserved true values for unit $a$ in sample $A$ and unit $b$ in
sample~$B$. We assume that, conditionally on their respective true
values and a parameter vector $\beta=(\beta_1,\ldots,\beta_h)$ which
accounts for the measurement error, $x^A$~and~$\xB$ are independent,
that is,
\[
p(x^A,x^B|\mu^A,\mu^B,\beta)=p(x^A|\mu^A,\beta) p(x^B|\mu^B,\beta);
\]
we also assume that, in each sample, all the observations
are conditionally independent given their true values and $\beta$. Then
\[
p(x^A|\mu^A,\beta)=\prod_{a=1}^{n_A} p(x^A_a|\mu^A_a,\beta),\qquad
p(x^B|\mu^B,\beta)= \prod_{b=1}^{n^B} p(x^B_b|\mu^B_b,\beta),
\]
with
\[
p(x^A_a|\mu^A_a,\beta)=\prod_{i=1}^h p(x_a^{A_i}|\mu_a^{A_i},\beta_i),\qquad
p(x^B_b|\mu^B_b,\beta)=\prod_{i=1}^h p(x_b^{B_i}|\mu_b^{B_i},\beta_i).
\]
Note that the vectors $\mu^A$ and $\mu^B$ introduce a first latent
structure into our record linkage model and make it effectively a
missing data model [\citet{mvf09}].

We conclude the top stage of the hierarchical structure by explicitly
introducing the measurement error model. A general model for
potentially misclassified observed records can be formulated as
$p(x^i=v^i_{j_i}|\mu^i=v^i_{j'_i})$, for all $(j_i,j'_i)$. Such a
model has been considered, in a Bayesian framework, by
\citet{swartzetal} who discuss several identifiability problems, and
by \citet{peirezetal}, where strong prior information is introduced in
the model. Here, to maintain the number of parameters in the model
reasonably low, we propose a simpler version of the so-called
\textit{hit--miss} model [\citet{copashilton}]
\[
p(x^{i}=v^i_{j_i}|\mu^{i}=v^i_{j_i'})=\beta_i
I({v^i_{j_i}=v^i_{j_i'}})+(1-\beta_i)/k_{i},\qquad  i=1,\ldots, h,
\]
where $\beta_i$ represents the probability of observing the true value
for the $i$th variable ``not by chance'' and $k_i$ is the number
of levels of variable $x^i$. This way, conditionally on the
unobserved true values, each single record field can be modeled as a
mixture of two components: the first component is concentrated on the
true value, while the second one is uniformly distributed over the set
$v^i=\{v^i_1, \ldots,v^i_{k_i}\}$. For a recent implementation of the
hit--miss model see also \citet{noren05}.

We now specify the conditional distributions of $\mu^A$ and $\mu^B$.
In particular, we assume that $\mu^A$ and $\mu^B$ are two independent
simple random samples drawn without replacement from a finite
population of unknown size $N$. The unknown vector
$F=(F_1,\ldots,F_j,\ldots,F_k)$, $k=\prod_{i=1}^h k_i$, represents the
population counts for each element $v_{j}$ of the set $V$. Obviously,
$\sum_{j=1}^k F_j=N$. In principle, one can write the model for the
unobserved true values $\mu^A$ and~$\mu^B$ in the following natural
way:
%
\begin{equation}
\label{mu.indipendenza} p(\mu^A,\mu^B|F)=p(\mu^A|F) p(\mu^B|F)
\end{equation}
with
%
\begin{equation}
\label{mu.marginale} p(\mu^S|F)=\pmatrix{n^S\cr f_1^S,\ldots,f^S_k}^{-1}
\left[\pmatrix{N\cr n^S}^{-1} {\prod_{j=1}^k
\binom{F_j}{f_j^S}} \right]\qquad  S=A,B,
\end{equation}
where $f^S=(f_1^S,\ldots,f_j^S,\ldots,f_k^S)$, $S=A,B$,
are the true sample counts (which are, however, unobservable, due to
measurement error) for each element $v_{j}\in V$. Formula
(\ref{mu.marginale}) can be obtained by noticing that the observed
values of $\mu^S$ determine the frequencies $f^S$, so that
$p(\mu^S|F)=p(\mu^S|f^S,F) p(f^S|F)$ where $p(\mu^S|f^S,F)$ and
$p(f^S|F)$ correspond to the two terms in (\ref{mu.marginale}). The
usual constraints $0\leq f_j^S\leq F_j$, $S=A,B$, must hold.

An alternative way of writing the above model is based on the use of
two latent quantities, which will play a crucial role in our approach.
The first quantity is the so-called matching matrix $C$. This is a
$n^A\times n^B$ matrix whose generic element $C_{ab}$ is a Bernoulli
random variable indicating whether or not unit $a$ in sample $A$ and
unit $b$ in sample $B$ are the same unit, that is,
\[
C_{ab}= \cases{
1,&\quad  if $ (a,b) \in M$, \cr
0,& \quad if $ (a,b) \in U$.
}
\]
The matrix $C$ is the actual quantity of interest in record linkage
problems; a similar structure also appears in different statistical
problems, such as the Bayesian alignment [\citet{greenmardia}] or
microarrays analysis [\citet{domu}]. We assume that multiple matches
are not possible. This implies that $ \sum_a C_{a b}\leq1\
\forall b=1,\ldots,n^B $, $ \sum_b C_{a b}\leq1\ \forall
a=1,\ldots, n^A $; also, note that there are $\binom{n^A}{T}
\binom{n^B}{T} T!$ different $C$ matrices with exactly $T=\sum_{a b}
C_{a b}$ matches, $T\leq\min(n^A, n^B)$.

The other latent quantity we introduce is the vector
$t=(t_1,\dots,t_j,\ldots,t_k)$ denoting, for each element of $V$, the
number of matches having $v_{j}$ as the true value. The vector $t$
(which is basically needed to facilitate the simulation of the
posterior distribution, as outlined in the following section) is a~%
deterministic function of $\mu^A,\mu^B$ and $C$.

Consider, as an illustration, the case where $\mu$ is univariate and
$V=\{v_1,v_2,\break v_3,v_4\}$: suppose we have $\mu^A=(v_1,v_2,v_1)$,
$\mu^B=(v_2,v_3,v_1,v_2)$, with $C_{1 3}=C_{2 4}=1$ and all the
other elements of $C$ equal to 0; then $t=(1,1,0,0)$. Finally, notice
that $0\leq t_j\leq\min \{f_j^A,f_j^B \}$ $\forall
j=1,\ldots,k$ and $\sum_{j=1}^k t_j=T$.

Now we introduce the model assumptions for the conditional
distribution of $\mu^A$ and $\mu^B$ given the values of $t,C$ and
$F$. First, note that $p(\mu^A,\mu^B|t,F,\break C)=0$ when $\mu^A_a \neq
\mu^B_b$ and $C_{ab}=1$. Also, we have $p(\mu^A,\mu^B|t,F,C)=0$
either when $\min\{f_j^A,f_j^B \} <t_j$ or
$\max\{f_j^A,f_j^B \} >F_j$. In any other situation it
turns out that
%
\begin{eqnarray}
\label{condmu1} p(\mu^A,\mu^B|C,t,F)&=& \frac{\prod_{j=1}^k
\binom{F_j-t_j}{f_j^A-t_j,f_j^B-t_j,F_j-f_j^A-f_j^B+t_j}}
{\binom{N-T}{n^A-T,n^B-T,N-n^A-n^B+T}} \nonumber\\[-8pt]\\[-8pt]
&&{}\times \frac{\prod_{j=1}^k
t_j! (f^A_j-t_j)! (f_j^B-t_j)!}{T! (n^A-T)! (n^B-T)!}. \nonumber
\end{eqnarray}
The distribution in (\ref{condmu1}) has the following
interpretation: the first term is the joint distribution of the sample
counts $f^A$ and $f^B$, say, $p(f^A,f^B|C,t,F)$; it can be obtained by
observing that, given the vector $t$, there are already~$t_j$ elements
in the category $v_j$, $j=1,\ldots, k$. Then, out of the total number
of partitions of the $N-T$ elements actually sampled in three disjoint
sets\footnote{They respectively represent the ``nonmatch'' for
samples $A$ and $B$ and the ``nonsampled'' units.} of sizes $n^A-T$,
$n^B-T$ and $ N-n^A-n^B+T$, one should only consider those where
category $v_j$ respectively appears $f_j^A-t_j$, $f_j^B-t_j$ and
$F_j-f_j^A-f_j^B+t_j$ times in the three sets, for $j=1, \ldots, k$.
The other term in (\ref{condmu1}) is the conditional distribution
$p(\mu^A,\mu^B| f^A,f^B,C,t,F)$; given $f^A$ and $f^B$, the matching
matrix $C$ and the vector $t$, there are
\[
T! (n^A-T)! (n^B-T)!
\]
possible permutations of the elements of the two samples: among them,
there are $\prod_j t_j ! ( f_j^A-t_j)! (f_j^B-t_j)!$ permutations
which exactly reproduce the orderings given in $\mu^A$ and $\mu^B$.

The prior distribution for $C$ and $t$ should reflect the random
selection mechanism of the two samples. Conditionally on $t$ and $F$,
$C$ has a uniform distribution on the set of all possible matching
matrices with $T$ matches. Loosely speaking, in the absence of
information about $\mu^A$ and $\mu^B$, all the possible couples are
equally likely to be a match. Then, we have $p(C,t|F)=p(C|t,F)p(t|F)$
with
\[
p(C|t,F)=p(C|t)= \cases{
0, &\quad  if $\displaystyle \sum_{ab}C_{ab}\neq\sum_j
t_j$,\cr
\displaystyle \left[ \pmatrix{n^A\cr T} \pmatrix{n^B\cr T} T!
\right]^{-1},
&\quad otherwise.
}
\]
To derive the distribution $p(t|F)$, one can
observe that, given $T$, $t$ is a~vector of counts of the $v_j$
categories in a simple random sample of size $T$ drawn from the
population. Then $p(t|T,F)$ is a multivariate hypergeometric
distribution. Finally, $T$, the total number of common units across
the two samples, is a~scalar hypergeometric random variable. Then,
%
\begin{eqnarray}
\label{t.f}
p(t|F)&=&p(t|T,F)p(T|F)\nonumber\\[-8pt]\\[-8pt]
&=&
\prod_{j=1}^k \left[ \pmatrix{F_j\cr t_j}\Big/{\pmatrix{N\cr T}} \right]
{\pmatrix{n^A\cr T} \pmatrix{N-n^A\cr n^B-T}}\Big/{\pmatrix{N\cr n^B}}.\nonumber
\end{eqnarray}
It is easy to see that, by averaging out over $C$ and
$t$ in the distribution $p(\mu^A,\mu^B,C,t|F)$, one re-obtains the
model expressed by (\ref{mu.indipendenza}) and
(\ref{mu.marginale}). Details are given in Appendix \ref{appendixA}. For the moment
notice that the use of the hypergeometric distribution $p(T|F)$ in
(\ref{t.f}) is standard practice in capture--recapture modeling when
the number $T$ of common units across two samples is known
[\citet{darroch}, \citet{seber86} and \citet{marinrobert}].

At the bottom of the hierarchical model, one needs to specify the
prior for the vector $F$; this is equivalent to assuming that the
finite population which the two samples are drawn from is itself a
random sample from a~superpopulation model [\citet{ericson69}]. In
particular, following \citet{hoadly}, we assume that, conditionally on
$N$ and a vector $\theta=(\theta_1,\ldots,\theta_k)$, with $0\leq
\theta_i\leq1$ and $\sum_{i=1}^k\theta_i=1$, $F$ is a multinomial
random variable,
\[
p(F_1,\ldots,F_k|\theta,N)=
\frac{N!}{F_1!F_2!\cdots F_k!} \prod_{j=1}^k \theta_j^{F_j}.
\]
Regarding
the prior for $N$, we suggest the following family of noninformative
priors:
\[
p_g(N)\propto\Gamma(N-g+1)/N!,\qquad  g\geq0;
\]
the
hyperparameter $g$ regulates the shape of the prior: the larger the
value of $g$, the lower the prior weight on the right tail, which is
integrable for all $g>1$. The same prior model for $F$ can be
expressed by assuming that, for a fixed hyperparameter $\lambda>0$ and
$\theta$, the population counts $F_1,\ldots, F_k$ are independent
Poisson variables with rates $\lambda\theta_1,\ldots,\lambda\theta_k$
and $p_g(\lambda) \propto1/\lambda^g$.

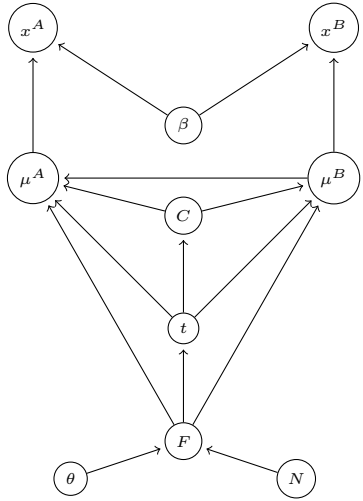
\begin{figure}
\begin{center}
\begin{tiny}
\begin{tikzpicture}
\begin{scope}[shape=circle]
\tikzstyle{every node}=[draw]
\node(xa) at (-2,1) {$x^A$};
\node(xb) at (2,1) {$x^B$};
\node(mua) at (-2,-1) {$\mu^A$};
\node(mub) at (2,-1) {$\mu^B$};
\node(F) at (0,-4.5) {$F$};
\node(t) at (0,-3) {$t$};
\node(C) at (0,-1.5) {$C$};
\node(theta) at (-1.5,-5) {$\theta$};
\node(N) at (1.5,-5) {$N$};
\node(beta) at (0,-0.3) {$\beta$};
\end{scope}
\draw[->,shorten >=2pt] (F) -- (mua);
\draw[->,shorten >=2pt] (F) -- (mub);
\draw[->,shorten >=2pt] (F) -- (t);
\draw[->,shorten >=2pt] (theta) -- (F);
\draw[->,shorten >=2pt] (t) -- (mua);
\draw[->,shorten >=2pt] (t) -- (mub);
\draw[->,shorten >=2pt] (N) -- (F);
\draw[->,shorten >=2pt] (t) -- (C);
\draw[->,shorten >=2pt] (C) -- (mua);
\draw[->,shorten >=2pt] (C) -- (mub);
\draw[->,shorten >=2pt] (mua) -- (xa);
\draw[->,shorten >=2pt] (mub) -- (xb);
\draw[->,shorten >=2pt] (mub) -- (mua);
\draw[->,shorten >=2pt] (beta) -- (xa);
\draw[->,shorten >=2pt] (beta) -- (xb);
\end{tikzpicture}
\end{tiny}
\end{center}
\caption{DAG representation of the joint probability model described in Section
\protect\ref{model}.}\label{grafo}
\vspace*{8pt}
\end{figure}

Last, we assume that the prior for $\theta$ is obtained first by
modeling its elements via the product of marginal and conditional
probabilities based on a~specific association pattern for the key
variables and then by putting independent Dirichlet distributions to each
probability vector characterizing the resulting model for $\theta$. A~special case of this product of Dirichlet distributions is the
hyper-Dirichlet prior which is used in the similar context of
disclosure risk assessment by \citet{forsterwebb}; see also
\citet{ohaganforster}. Moreover, the ``measurement error'' parameters
$\beta$ are independent and uniformly distributed random variables;
they are also independent of all the other model parameters.
To sum up, the joint distribution of all the variables is expressed
by the following factorization:
\begin{eqnarray*}
p(x^A,x^B,\mu^A,\mu^B,\beta,C,t,F,N,\theta)&=&
p(x^A,x^B|\mu^A,\mu^B,\beta)\\
&&{}\times p(\mu^A,\mu^B|F,C,t)p(C|t)p(t|F)\\
&&{}\times p(F|\theta,N)p(N)p(\theta)p(\beta),
\end{eqnarray*}
and a representation in terms of a directed acyclic
graph is displayed in Figure \ref{grafo}.

\section{Bayesian implementation}
\label{MCMC}

In this section we discuss a \textit{Metropolis within Gibbs}
algorithm for simulating from the joint posterior distribution
$p(\mu^A, \mu^B,\break \beta, t,F,N,\theta\mid x^A, x^B)$; see
\citet{robertcasella04} for a general overview about the MCMC theory
and implementation. In our case the Gibbs algorithm structure is based
on the following updating steps:\vspace*{2pt}
\begin{eqnarray*} \mu^A,\mu^B,t
&\hspace*{-2pt}|\hspace*{-2pt}&F,N,\theta,\beta,\\[2pt]
F,N&\hspace*{-2pt}|\hspace*{-2pt}&\mu^A,\mu^B,t,\theta,\beta,\\ \theta&\hspace*{-2pt}|\hspace*{-2pt}&\mu^A,\mu
^B,t,F,N,\beta,\\[2pt]
\beta&\hspace*{-2pt}|\hspace*{-2pt}&\mu^A,\mu^B,t,F,N,\theta.
\vspace*{2pt}
\end{eqnarray*}
When the matching matrix $C$ is itself one of the
parameters of interest, one can simply add, at each iteration of the
algorithm, a draw from the conditional distribution\vspace*{2pt}
\[
p(C|\mu^A \mu^B, \beta,t,F,N,\theta,x^A,x^B).\vspace*{2pt}
\]
Details about this conditional distribution are given later in this
sec-\break tion.

To illustrate the first updating step, notice that\vspace*{2pt}
\begin{eqnarray*} p(\mu^A,\mu^B,t |F,\theta,\beta,x^A,x^B)&=&
p(\mu^A|F,\beta,x^A) p(\mu^B|F,\beta,x^B)\\[2pt] &&{}\times p(t|
\mu^A,\mu^B,F,\theta,\beta,x^A,x^B).\vspace*{2pt}
\end{eqnarray*}
Moreover, by using results from Appendix \ref{appendixB},
%
\begin{eqnarray}
\label{tdasola}
p(t| \mu^A,\mu^B,F,\theta,\beta,x^A,x^B)& =& p(t
|\mu^A,\mu^B,F)
\nonumber
\\[-6pt]
\\[-6pt]
\nonumber
&=&\prod_{j=1}^k \left[ \frac{\binom{f_j^A}{t_j}
\binom{F_j-f_j^A}{f_j^B-t_j}}{\binom{F_j}{f_j^B}} \right] 
\end{eqnarray}
Thus, conditionally on all the other quantities, $t_1,
\ldots, t_k$ are independent hypergeometric random variables. Then
one should separately draw $\mu^A$ and $\mu^B$ from
$p(\mu^A|F,\beta,x^A)$ and $p(\mu^B|F,\beta,x^B)$ and $t$ from
(\ref{tdasola}). However, the direct simulation of $\mu^A$ and
$\mu^B$ is not straightforward. To see why, let
$F_{\mu|\mu_1,\ldots,\mu_l}$ be the population count for the category
assumed by $\mu$ after eliminating, from the population, $l$ units
with categories $\mu_1,\ldots,\mu_l$. Then
\begin{eqnarray*} p(\mu^S|F,\beta,x^S)&\propto&
p(\mu^S|F)p(x^S|\mu^S,\beta) \\[2pt]
&\propto& \prod_{s=1}^{n^S}
F_{\mu^S_s|\mu^S_1,\ldots,\mu^S_{s-1}} \prod_{i=1}^k \biggl[\beta_i
I_{\{\mu_s^{S_i}=x^{S_i}_s\}}+(1-\beta_i) \frac{1}{k_i} \biggr]
\end{eqnarray*}
for $S=A,B$ and the direct simulation from the above
distributions can be computationally hard. To circumvent\vadjust{\goodbreak} the
difficulty of directly simulating the entire joint distribution
$p(\mu^S|F,\beta,x^S)$, note that we can easily draw the full
conditionals
%
\begin{eqnarray} \label{mucondgibbs}
p(\mu^S_s|\mu^S_{-s},\beta,x^S)&\propto&
F_{\mu^S_s|\mu^S_1,\ldots,\mu^S_{s-1},\mu^S_{s+1},\ldots,\mu^S_{n^S}}\nonumber\\[-6pt]\\[-6pt]
&&{}\times\prod_{i=1}^k \biggl[\beta_i I_{\{\mu_s^{S_i}=x^{S_i}_s\}}+(1-\beta_i)
\frac{1}{k_i} \biggr]\nonumber
\end{eqnarray}
for $s=1,\ldots,n^S$ and $S=A,B$. By simulating $\mu^A$
and $\mu^B$ from (\ref{mucondgibbs}) following a Gibbs type updating
and $t$ by its true conditional distribution, we do not produce an
exact draw from the conditional distribution of $(\mu^A,\mu^B,t)$.
However, the latter is exactly the stationary distribution associated
with the proposed step. This strategy can then be justified as an
example of ``Metropolis within Gibbs.'' Moreover, note that, in order
to improve the mixing of the chain, for each iteration of the algorithm
we can repeat more simulation cycles from the conditional distributions
(\ref{mucondgibbs}) in order to approximately generate, at each
iteration, a random draw from the true conditional of
$(\mu^A,\mu^B,t)$.

A standard Gibbs updating is possible for the second step. Consider
the full conditional distribution of the vector $F$; using the results in
Appendix~\ref{appendixA} and after some algebra,
\begin{eqnarray*} p(F|\mu^A,\mu^B,t,\theta,\beta,x^A,x^B)&\propto&
p(\mu^A,\mu^B|F,t)p(t|F)p(F|\theta)\\ \nonumber&\propto&
\prod_{j=1}^k \frac{F_j!}{(F_j-f_j^B-f_j^A+t_j)!}
\frac{\theta_j^{F_j}}{F_j!} \frac{\Gamma(N-g+1)}{ \binom{N}{n^A}
\binom{N}{n^B}}\\ \nonumber&\propto& (N-n^A-n^B+T)! \prod_{j=1}^k
\frac{\theta_j^{F_j-f_j^B-f_j^A+t_j}}{(F_j-f_j^A-f_j^B+t_j)!}
\\\nonumber&&{}\times \frac{\Gamma(N-g+1)}{ (N-n^A-n^B+T)!
\binom{N}{n^A} \binom{N}{n^B}}.
\end{eqnarray*}
Then, random draws from the above distribution can
easily be obtained by first simulating $N$ from
%
\begin{equation}
\label{condN} p(N|T)\!\propto\frac{\Gamma(N-g+1)}{ (N-n^A-n^B+T)!
\binom{N}{n^A} \binom{N}{n^B}}\!\propto\frac{ \binom{n^A}{T}
\binom{N-n^A}{n^B-T}}{ \binom{N}{n^B}}\frac{\Gamma(N-g+1)}{N!}.\hspace*{-30pt}
\end{equation}
Subsequently, conditionally on $N$, one can draw
$v_1,\ldots,v_k$ from a multinomial distribution with parameters
$\theta_1,\ldots,\theta_k$ and size $N-n^A-n^B+T$, and then set
$F_j=v_j+f_j^A+f_j^B-t_j$.

Incidentally, we notice that the posterior distribution (\ref{condN})
plays a crucial role also when the sample sizes $n^A$ and $n^B$ are
assumed to be random and $T$ is known. In fact, in this case, the
vector $[T, n^A-T, n^B-T, N-n^A-n^B+T]$ follows a multinomial
distribution with parameters $N$ and $(p^A p^B, p^A(1-p^B),
p^B(1-p^A), (1-p^A)(1-p^B))$, where $p^A$ and $p^B$ represent the
unknown capture probabilities in the two sampling occasions; see, for
example, \citet{bishop}. It follows that, for $(p^A, p^B)$ unknown,
inference for $N$ can be drawn either by using the complete model
[i.e., by introducing a prior for $(p^A, p^B)$ and then getting the
marginal posterior distribution $p(N|n^A,n^B,T)$] or, in a slightly
approximate way, by eliminating $(p^A,p^B)$ via a conditional argument
[i.e., by using the conditional likelihood $p(T|n^A,n^B,N)$]. These
two approaches typically produce very similar conclusions. In the
former case, when assuming a uniform prior for $(p^A, p^B)$, the
marginal posterior of $N$ is given by the expression for $p(N|T)$ in
(\ref{condN}) multiplied by $(N+1)^{-2}$. In the latter case,
inference is based only on (\ref{condN}). The complete multinomial
likelihood can obviously be used within our approach by simply adding
other Gibbs steps for $(p^A,p^B)$. However, as in the case with known
$T$, we do not expect to see substantial differences, and in the rest
of the paper we will consider $n^A$ and $n^B$ as fixed.

The updating of $\theta$ can be done in a standard way since
\[
p(\theta|\mu^A,\mu^B,t,F,N,\beta,x^A,x^B)\propto p(\theta)
p(F|\theta)
\]
and
the independent Dirichlet distributions characterizing $p(\theta)$ are
conjugate to $p(F|\theta)$; see \citet{ohaganforster}. Finally, note
that the conditional posterior density for $\beta_i$ is proportional
to
\[
\bigl(\beta_i+(1-\beta_i)/k_i\bigr)^{\tilde{n}^{AB}_i}(1-\beta_i)^{n^A+n^B-
\tilde{n}^{AB}_i},
\]
where $\tilde{n}^{AB}_i$ is the total number of
sample units where the observed value and the true value coincide for
the $i$th key variable. One can easily see that the posterior
distribution of $\eta_i=\beta_i+(1-\beta_i)/k_i$, conditionally on all
the other variables, is
Beta($\tilde{n}^{AB}_i+1,n^A+n^B-\tilde{n}^{AB}_i+1$) truncated on
the set $(k_i^{-1},1)$. Then we draw $\eta_i$ from its Beta
distribution and set $\beta_i=(k_i\eta_i-1)/(k_i-1)$, for
$i=1,\ldots,k$.

\subsection{Matching matrix simulation}

In order to specify the conditional distribution of $C$ given all
other quantities involved in the model, we introduce the sets
$A_j= \{a\dvtx \mu^A_a=v_j \}$ and $B_j= \{b:
\mu^B_b=v_j \}$. In words, $A_j$ is the set of units in sample
$A$ whose true value belongs to category $v_j$; these sets depend on
$\mu^A$ and $\mu^B$. Let $C_j$ be the block of the matrix $C$
corresponding to the rows in $A_j$ and the columns in
$B_j$. Conditional on the true values, $\mu^A$ and $\mu^B$, $C_{ab}=0$
for each couple such that $\mu^A_a\neq\mu^B_b$; then, outside the
blocks $C_1,\ldots, C_k$, the elements of $C$ will be equal to
0. Thus,
\[
p(C|\mu^A,\mu^B,t,F,\theta,x^A,x^B)=\prod_{j=1}^k
p(C_j|t_j, f_j^A,f_j^B),
\]
where $p(C_j|t_j,f_j^A,f_j^B)$ is the
discrete uniform distribution over the set of all possible
configurations for the block $C_j$ with exactly $t_j$ matches,
\[
p(C_j|t_j,f_j^A,f_j^B)= \left[
{ t_j! \pmatrix{f_j^A\cr t_j} \pmatrix{f_j^B\cr t_j}} \right]^{-1}.
\]

Note that, by conditioning on the drawn values of the key variables,
we automatically create a blocking method able to limit the number of
candidate matches. Blocking strategies are very popular in record
linkage literature. They basically consist of a partition into
homogeneous groups of all the possible comparisons among records in
order to reduce the computational burden; see, for example,
\citet{newcombe67} or \citet{winkler04}. Within our approach the
homogenous groups of records are identified at each step of the
algorithm by the block matrices $C_j$'s.

\subsection{Matching matrix estimation via MCMC algorithm}

Now we describe inferential strategies for producing a ``point
estimate'' in a record linkage analysis. The usual output of an MCMC
based analysis is a sample of approximately independent
``observations,'' simulated from the posterior distribution. This
sample can be used to obtain a representation of the uncertainty about
the parameters of interest, mainly the matrix $C$ or $N$. In
addition, record linkage procedures are often the first stage of a
more complex statistical analysis: they represent the crucial step of
creating a suitable data set to be used afterward. In terms of
statistical theory, this is equivalent to producing a point estimate of
$C$, from which we select the ``declared'' matches. Classical
inference methods usually provide plug-in estimates, based on theories
developed in \citet{fesu69} and \citet{jaro89}. First, the
previously defined parameters $m$ and $u$ are estimated and then a
sequence of statistical tests is performed in order to decide whether
each pair $(a,b)\in A\times B$ can be declared a match or not. The
power of multiple tests is calibrated in order to obtain a specific
level of the False Match Rate (FMR), that is, the ratio between
the number of false matches and the total number of declared matches.
Note that the FMR is exactly equivalent to the well-known False
Discovery Rate [\citet{benhoc95}], very popular in multiple comparison
applications (wavelets theory, microarray analysis, etc.) Furthermore,
currently used record linkage procedures must complete the statistical
data analysis with a reallocation procedure which eliminates
inconsistencies among the results of different tests [see
\citet{jaro89} and the problem posed by \citet{lar99}, paragraph 3.3].

The Bayesian way of facing a record linkage problem is different in
spirit, and suggests interesting issues, both from a practical and a
methodological perspective. Although in a formal Bayesian analysis
one should select the point estimate as the one minimizing the
posterior expected loss, it is common practice, in applications, to
use the posterior mean or, sometimes, the posterior median. Of
course, these solutions do not appear reasonable in a record linkage
context: the marginal posterior mean of each single element of the
matrix $C$ will be a number between 0 and 1, which does not help much
in deciding whether the pair $(a,b)$ is a match or not. The use of the
posterior median is even more complicated in multivariate discrete
settings. Thus, a formal decision theoretic approach seems necessary:
let $G=\{G_{ab}\}\in{\mathcal G}, a=1,\ldots,n^A$ and $b=1, \ldots,
n^B$, a generic matrix of size $n^A \times n^B$, with the same
characteristics as $C$, such that it represents our ``action.'' Here
${\mathcal G}$ represents the set of all possible actions. Also, let
$L(\cdot,\cdot)$ be a loss function defined as $L: {\mathcal G}\times
{\mathcal C} \rightarrow R^+$ where ${\mathcal C}$ is the set of all
possible matching matrices. Our goal is to select, for a given loss
function $L$, the optimal decision $G^\ast$, the one which minimizes
the posterior expected loss
\[
G^* = \argmin\limits_{G\in{\mathcal G}} W(G)
\]
where $W(G) = E[L(C,G)|x^A,
x^B]$. In what follows we will consider some specific loss functions:
\begin{itemize}[(3)]
\item[(1)] Quadratic Loss
\[
L_q(C,G)=\sum_a \sum_b (\cab-G_{ab})2.
\]
Since the elements of $C$ and $G$ are either 0 or 1, $L_q$ is equivalent
to the~$L_1$ loss: $L_1(C,G)=\sum_a \sum_b |\cab-G_{ab}|.$
\item[(2)] False Match Rate
\[
L_{\mathrm{FMR}}(C,G)=
\cases{
0, &\quad  if $\displaystyle \sum_a \sum_b G_{ab}= 0$, \cr
\displaystyle \frac{\sum_a \sum_b G_{ab} I({\cab=0})}{\sum_a \sum_b G_{ab} }, & \quad
otherwise.
}
\]
$L_{\mathrm{FMR}}$ translates, in terms of decision theory, the
classical use of the False Match Rate as a measure of performance of
the record linkage analysis.
\item[(3)] Absolute number of errors
\[
L_{\mathrm{ABS}}(C,G)= \sum_a \sum_b [ G_{ab} I({\cab=0}) + (1-G_{ab})
I({\cab
=1}) ].
\]
\end{itemize}
The following theorem provides the optimal solution for
the above mentioned losses.
\begin{thm}
\label{lossf}
\begin{itemize}[(A)]
\item[(A)] Under losses $L_q$ and $L_{\mathrm{ABS}}$, the optimal Bayesian
solution is given by the matrix $G^\ast$, defined as
\begin{eqnarray}
G_{ab}^\ast=
\cases{
1, & \quad  if $ p(\cab=1| x^A,x^B)>\frac12$,
\cr
0, &\quad  otherwise,
}\nonumber \\
\eqntext{a=1, \dots, n^A; b= 1,\dots, n^B.}
\end{eqnarray}
\item[(B)] Under loss $L_{\mathrm{FMR}}$, the optimal solution is a matrix
consisting of all zeros.
\end{itemize}
\end{thm}
\begin{pf}
First, notice that $I(\cab=1)=\cab$ and $I(\cab=0)=1-
\cab$.

(A): Since
\[
L_q(C,G) = \sum_a \sum_b [ \cab+ G_{ab} - 2 \cab G_{ab}
],
\]
the problem is equivalent to the maximization of the posterior
expected value of
\[
{L}_q(C,G) = 2 \sum_a \sum_b G_{ab} \biggl[ \cab-\frac12
\biggr].
\]
With the loss $L_{\mathrm{ABS}}$, simple calculations lead to
\begin{eqnarray*} L_{\mathrm{ABS}}(C,G) &=& \sum_a \sum_b [ G_{ab}
(1 - \cab) + ( 1 - G_{ab} ) \cab]\\ &=&
\sum_a \sum_b [ G_{ab} - 2 G_{ab} \cab+ \cab].
\end{eqnarray*}
The minimization of the posterior expected loss of
$L_{\mathrm{ABS}}$ is equivalent to the maximization of the quantity
\[
{L}_q(C,G) = 2 \sum_a \sum_b G_{ab} \biggl[\cab-\frac12
\biggr].
\]
Then the quantities ${L}_q$ and ${L}_{\mathrm{ABS}}$
are identical and it will be sufficient to find the optimal solution
for ${L}_q$. We need to maximize
\begin{eqnarray*} {W}_q(G)&=& 2 E \biggl(\sum_a \sum_b
G_{ab} \biggl[ \cab-\frac12 \biggr] \Big| x^A,x^B
\biggr)\\ &=& 2 \sum_a \sum_b G_{ab} \biggl[ p(\cab=1|x^A,x^B
)-\frac12 \biggr].
\end{eqnarray*}
The last expression shows that the value that
maximizes ${W}_q(G)$ is obtained by setting $G_{ab}=1$ if
and only if the correspondent coefficient is positive, that is, when
$p(\cab=1|x^A,x^B ) >\frac12$.

(B): When $L_{\mathrm{FMR}}$ is
used, it is easy to see that FMR is minimized by adopting the
conservative behavior of not declaring any match! In this case, in
fact, the posterior expected loss is always zero, independently of the
posterior distribution. Then the optimal solution is given by
$G_{a b}^\ast=0$, for all $(a,b)$.
\end{pf}

It is important to stress that all the optimal solutions derived in
Theorem~\ref{lossf} are based on the marginal posterior probabilities
of being a match for the various pairs $(a,b)$. This is a consequence
of the fact that the above loss functions are additive and they
basically ``sum'' over all the losses due to the single mismatches.

Part B of Theorem \ref{lossf} is also important. It says that, from a
decision theoretic perspective, the FMR is not a valid measure of
performance, because it only controls one type of error. Every
reasonable loss function should also take into account a measure of
the number of undiscovered matches [\citet{genowa01}]. In this sense,
a reasonable loss function for record linkage may be given by the
\textit{Global Error Rate}
\[
L_{\mathrm{TOT}}(C,G)= L_{\mathrm{FMR}}(C,G) + \frac{\sum_a \sum_b (1-G_{a b})
I_{\cab=1}(\cab)}{\sum_a \sum_b (1-G_{ab}) }.
\]
The loss
$L_{\mathrm{TOT}}$ is actually able to capture errors due to missing true
matches. However, the improvement is more theoretical than practical:
in fact, the denominator of the second factor is so much larger than
the denominator of $L_{\mathrm{FMR}}$ that the results obtained using
$L_{\mathrm{TOT}}$ should not be practically different from those derived under
loss $L_{\mathrm{FMR}}$.

\section{Illustrative application}\label{appl}

We illustrate our approach in detail with the real data
set already used in the \hyperref[Introduction]{Introduction}. The two files consist of
$n^A=34$ records from a single block of the last Italian census
population survey and $n^B=45$ records from the same block relative to
the post enumeration survey; more details can be found in
\citet{aft07}. Records in both files refer to foreign residents
only, which typically represent an example of an elusive population. For
each file, we take three key variables: $X_1$ represents the first two
consonants of the family name with 339 observed categories
(considering all blocks), $X_2$ represents the gender and $X_3$ is the
education level, with 17 categories. The total number of entries in
$V$ is $k=11\mbox{,}526$. The data and the programs [written in C and R, \citet{RRR}]
that have been used for this application are available in
the supplementary material [\citet{sptl}]. In practice, real
applications may have
more key variables, more blocks and larger sample sizes. However,
focusing on a small example allows us to illustrate better some details
of our methodology compared to the existing approaches.

The hyperparameter $g$ appearing in the prior distribution $p(N)$ has
been set equal to 2 in order to have a proper prior. The Dirichlet
distributions for $\theta$ are chosen so that, at the superpopulation
level, $X_1$ is independent of $(X_2, X_3)$. We also assume that all
the Dirichlet distributions are uniform in their supports.

\begin{figure}

\includegraphics{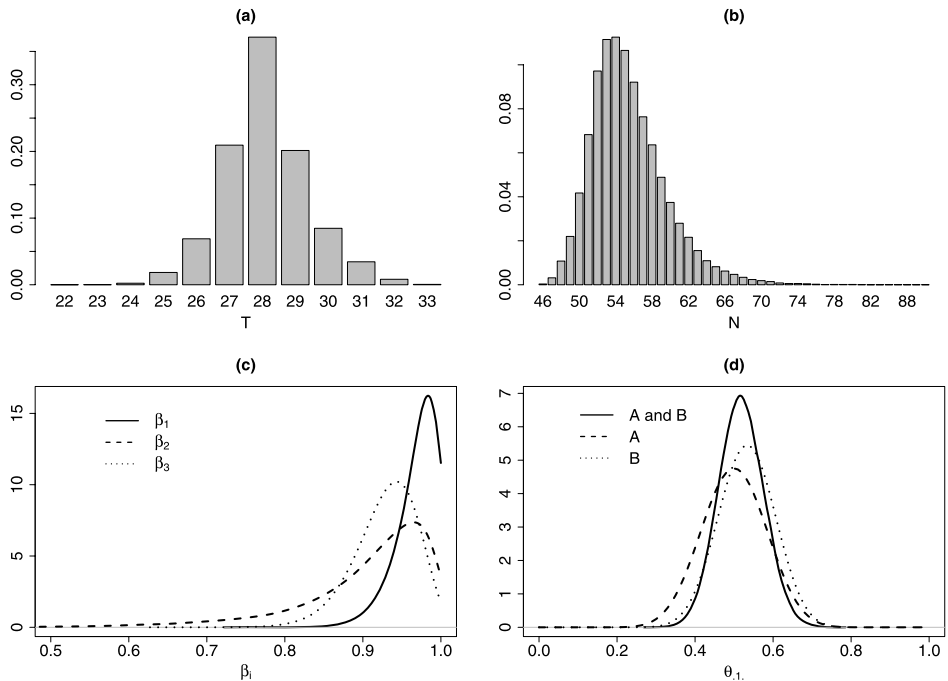}

\caption{Posterior distributions of the number of matches $T$, the
population size $N$, the parameter vector $\beta_i$ $(i=1,2,3)$ and
$\theta_{\cdot1 \cdot}$.}
\label{figura2}
\end{figure}

We have used the algorithm described in Section \ref{MCMC} to generate
a single Markov chain of length 100,000. See the supplementary material
for a~graphical representation of some of the simulation traces.
Figure \ref{figura2} shows the posterior distributions of the
following quantities: (a) the number of matches $T$,
(b) the total population size $N$, (c) the
measurement error parameter vector $\beta_i$, $(i=1,2, 3)$,
(d) the probability of selecting a male within the block at
the superpopulation level, $\theta_{\cdot1 \cdot}.$ In panel
(d), we also show the posterior distributions of $\theta_{\cdot1
\cdot}$ obtained by considering the two files separately, assuming a
uniform prior and independence among the units. Notice that the
posterior density of $\theta_{\cdot1 \cdot}$ can be graphically
interpreted as an average of the two posteriors one would have
obtained from the analysis of each single data set.

\begin{sidewaysfigure}

\includegraphics{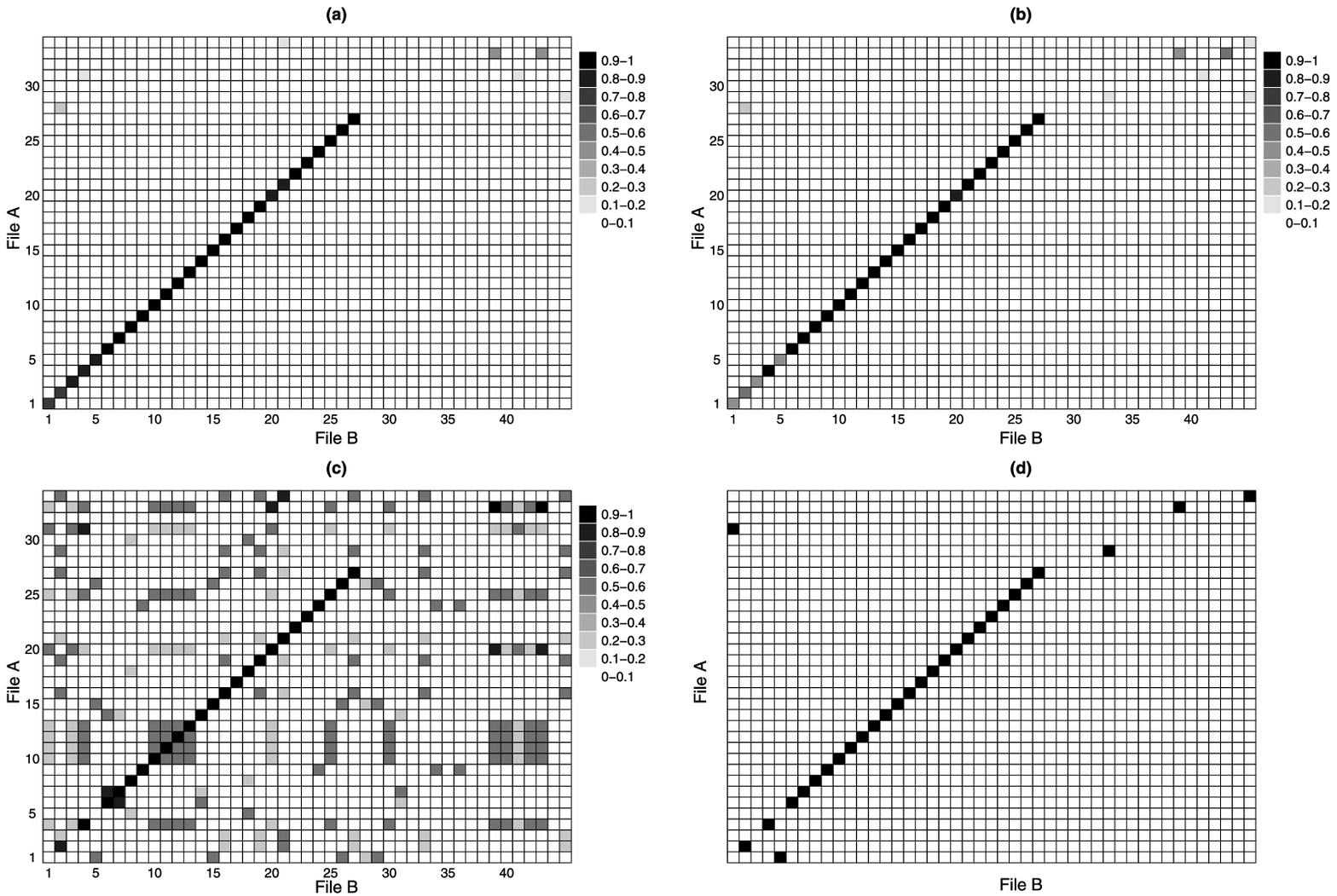}

\caption{Matching estimation. Panel \textup{(a)} shows the posterior
probabilities $p(C_{ab}=1|x^A,x^B)$ under the new model. Panel \textup{(b)}
shows the posterior probabilities $p(C_{a,b}=1|y_{11},\ldots
,y_{n^A,n^B})$ under the Jaro constrained model. Panels \textup{(c)} and \textup{(d)}
show the posterior probabilities $p(C_{ab}=1|y_{ab})$
and the estimated matching matrix using the classical approach. Values
of the posterior probabilities are indicated by the shading scale at
the right of each panel.}\label{Figmatch1}
\end{sidewaysfigure}

The posterior estimated quantiles of level $(0.05, 0.5, 0.975)$ for
$T$ are $(26, 28,\break 31)$. The same posterior summaries for $N$ are $(49,
55, 65)$. Marginal posterior probabilities of being a match,
$p(C_{ab}=1|x^A,x^B)$, are graphically displayed in panel (a) of Figure
\ref{Figmatch1}, where the cases have been sorted in order to have the
most probable matches on the diagonal. There are only 34 pairs of
records (out of 1530) such that $p(C_{ab}=1|x^A,x^B)$ is larger than
0.1. The estimated matching matrix, using the quadratic loss function
outlined in Section \ref{MCMC}, is given by the 27 matches visible on
the diagonal. Notice that inference about $C$ is quite robust with
respect to the choice of the hyperparameter $g$: when $g=1$ we
obtained exactly the same estimated matching matrix, while setting
$g=3$ would produce one more match.

\begin{figure}

\includegraphics{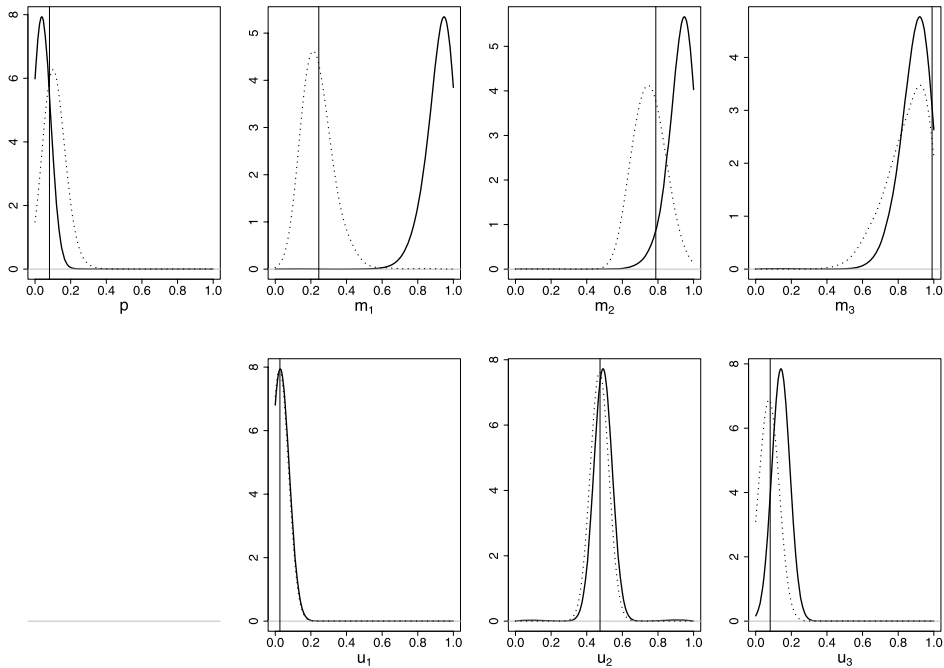}

\caption{Posterior distributions for the parameters of model
(\protect\ref{mixbern}) with the constraints on the matching matrix $C$ (solid
line) and without (dotted line). For the latter case the constraint
$p<1/2$ has been used to guarantee identifiability and the vertical
lines indicate maximum likelihood estimates.}\label{jarofig}
\vspace*{-6pt}
\end{figure}

We now compare our results with other possible approaches based on the
comparison vectors $y_{ab}$ whose frequency distribution is given in
Table~\ref{jarotab}. As a first alternative we consider a slight modification of
the Bayesian approach proposed by \citet{larsen05} where $y_{ab}$ is
marginally distributed as~(\ref{mixbern}) and the matching matrix $C$
satisfies the constraints $\sum_a C_{ab}\leq1$ and $\sum_b
C_{ab}\leq
1$. We use uniform priors for $m$ and $u$. Unlike \citet{larsen05}, we
have assumed, for the matching matrix $C$, the same prior distribution
used in our approach. We will call this model the ``Jaro constrained
model.'' The posterior distribution for the parameters $(m,u,C,N)$
can easily be simulated by using Gibbs steps for $[m|u,C,N]$,
$[u|m,C,N]$ and $[N|u,m,C]$. To update the matching matrix $C$, we use
the Metropolis--Hastings step proposed by \citet{greenmardia}. Figure
\ref{jarofig} reports the posterior distributions of the parameters
$p=T/(n^A\cdot n^B)$, $m$ and $u$. The posterior quantiles of level
$(0.05, 0.5, 0.975)$ for $T$ are estimated as $(23, 27, 31)$. The same
posterior summaries for $N$ are $(49, 57, 72)$. The marginal posterior
probabilities of being a match,
$p(C_{ab}=1|y_{11},\ldots,y_{n^A,n^B})$, are graphically displayed in
panel (b) of Figure \ref{Figmatch1}. Also in this case we have exactly
34 pairs of records such that $p(C_{ab}=1|y_{11},\ldots,y_{n^A n^B})$
is larger than 0.1, but the matching matrix obtained with the
quadratic loss provides 25 matches. In general, our proposed model and
the Jaro constrained model provide similar estimates, although the latter
seems to produce slightly more uncertainty as shown by the larger
interval estimates for both $T$ and $N$.

\begin{table}
\tablewidth=250pt
\caption{Results of the classic approach}\label{jarotab}
\begin{tabular*}{\tablewidth}{@{\extracolsep{\fill}}ld{3.0}cd{3.2}@{}}
\hline
$\bolds{y_{ab}}$ & \multicolumn{1}{c}{\textbf{Frequency}} & $\bolds{p((a,b) \in M| y_{ab})}$ & \multicolumn{1}{c@{}}{$\bolds{\lambda}$} \\
\hline
$(0,0,0)$ & 659 & 0.00 & 0.01 \\
$(1, 0, 0)$ & 20 & 0.01 & 0.14 \\
$(0, 1, 0)$ & 601 & 0.00 & 0.04 \\
$(1, 1, 0)$ & 13 & 0.05 & 0.58 \\
$(0, 0, 1)$ & 78 & 0.23 & 3.43 \\
$(1, 0, 1)$ & 8 & 0.80 & 45.20 \\
$(0, 1, 1)$ & 126 & 0.56 & 14.81 \\
$(1, 1, 1)$ & 25 & 0.94 & 194.97 \\
\hline
\end{tabular*}
\legend{The first
two columns
give the distribution of the comparison vector. The last two columns
report the estimated quantities
(\ref{jaropost}) and (\ref{FSratio}).}
\end{table}

Finally, we show the results obtained by considering model
(\ref{mixbern}) without row or column constraints on the matching
matrix $C$. Maximum likelihood estimates and posterior densities are
reported in Figure \ref{jarofig}. The matching step is performed by
considering the posterior matching probabilities (\ref{jaropost}) or
the likelihood ratios (\ref{FSratio}). In Table \ref{jarotab} we
report these quantities obtained with a simple plug-in of the maximum
likelihood estimates of the parameters. The posterior probabilities
$p((a,b)\in M|y_{ab})$ are also displayed graphically in panel (c) of
Figure \ref{Figmatch1}. In this case there are 237 pairs with a
posterior probability $p((a,b)\in M|y_{ab})$ greater than 0.1. The
higher number of {\em potential } matches is almost certainly due to
the fact that, in this approach, because of the independence
assumption among comparison vectors and the absence of constraints on
the $C$ matrix, the marginal matching probabilities only depend on the
information retrieved from the single comparison and not, as in the
previous models, on the information provided by the entire data set. To
rule out multiple matches, following \citet{jaro89}, we maximize the
function
%
\begin{equation}
\label{lp}\sum_{a=1}^{n^A} \sum_{b=1}^{n^B} z_{ab} \log{
\frac{\prod_{i=1}^k (\hat{m}^{y_{ab}^i}(1-\hat{m})^{1-y_{ab}^i} ) }
{\prod_{h=1}^k (\hat{u}^{y_{ab}^i}(1-\hat{u})^{1-y_{ab}^i} ) }}
\end{equation}
subject to the constraints $\sum_{a=1}^{\nu_A}
z_{ab}\leq1$ $\forall b$, $\sum_{b=1}^{\nu_B} z_{ab}\leq1$ $\forall
a$ and $z_{ab} \in\{0,1\}$ $\forall(a,b)$. The final answer produces
29 matches displayed in panel (d) of Figure~\ref{Figmatch1}. From
Table \ref{table1} one can see that, by setting $T=29$ in the hypergeometric
likelihood $ \binom{n^A}{T} \binom{N-n^A}{n^B-T}/\binom{N}{n^B}$ and
using the prior $p(N)\propto1/N^2$, one gets a 95\% credible interval
for $N$ equal to $[50,60]$, which is a subset of the intervals
obtained using our approach or the Jaro constrained model.

\section{Multiple block application}
\label{blocks}
In this section we illustrate the results obtained with a more
realistic exercise involving a multiple block scenario. In particular,
we repeated the analysis described in the previous section for each
census enumeration area (census block) also selected for the post
enumeration survey and including at least one foreign person during
the census survey. This way we obtained a list with 337 pairs of data
sets for a total of 3675 records taken on foreign people during the
2001 census population survey and 3404 analogous records originating
from the parallel post enumeration survey. The block sizes vary from a
minimum of one individual on at least one occasion to a maximum with
280 and 311 individuals on the two occasions. Note that the total
number of blocks selected for the post enumeration survey is 1098,
corresponding to 0.31\% of the total number of the Italian census
enumeration areas.

For each pair of data sets we performed a record linkage analysis in
order to estimate the total number of foreign people $N_{l}$ living in
the $l$th census block, for $l=1,\ldots,337$. In addition to
the three key variables considered in the single block analysis
outlined before, we also considered the age (coded into 10 categories).
At the superpopulation level in our hierarchical model we assumed the
surname to be independent of gender, education level and age. The
probability vector for the surname categories is assumed, as before,
to be uniform in its support. For the $340=2\times17 \times10$
joint probabilities of the other three key variables we set the
Dirichlet hyperparameters all equal to $1/340$ in order to avoid marginal
distributions that are too concentrated.

\begin{figure}[b]
\vspace*{-3pt}
\includegraphics{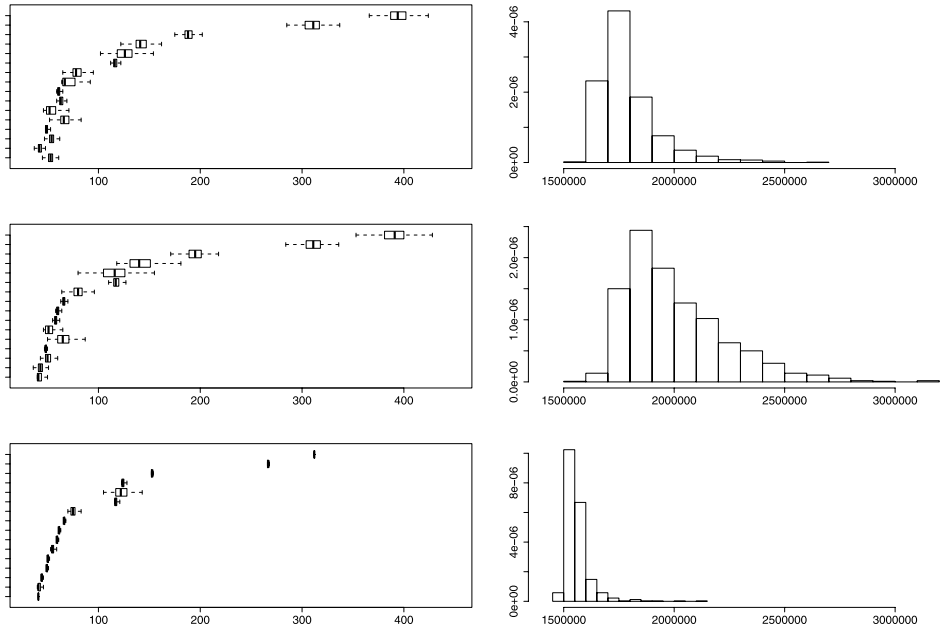}

\caption{\textup{Left panels}: box-plots from the posterior distribution of the
foreign population size for blocks with at least 25 matches.
\textup{Right panels}: posterior distribution of the foreign population size in
Italy at the end of 2001. \textup{Upper panels}: new model. \textup{Central panels}: Jaro
constrained model. \textup{Lower panels}: Jaro unconstrained hybrid approach.}\label{blockfigure}
\end{figure}

In the upper left panel of Figure \ref{blockfigure} we show, for each
block with approximately at least 25 matches, the box-plot for the
posterior distribution of~$N_{l}$ given by our approach. For each
pair of data sets we also implemented the other two approaches
described throughout the illustrative example, namely, the Jaro
constrained model and the hybrid strategy obtained by estimating the
matching matrix via the classical approach and then plugging in the
estimated match number in the posterior distribution of population
size. The box-plots for the posterior distribution of $N_{l}$ obtained
with these two approaches are shown, respectively, in the central and
lower left panels of Figure \ref{blockfigure}. Note that the posterior
distributions for $N_{l}$ provided by the Jaro constrained model give
point estimates similar to those obtained with our approach but with
slightly wider credibility intervals. Instead, the pattern shown by
the hybrid strategy is quite different. In particular, when compared
to the other approaches, it shows a remarkable under-estimation of the
block sizes. In fact, the maximization of the function (\ref{lp})
leads to an over-estimation of the true match number. However,
introducing a false match rate correction as in \citet{br95} would
reduce the distance from the other approaches. Nevertheless,
there is a clear message that ignoring the matching uncertainty would
give a false impression of accuracy for the estimates.

The same conclusions are emphasized when we aim at estimating the
quantity $N=[\sum_{l=1}^{337}N_{l}]/0.0031$ which can be seen as a
rough approximation for the size of the foreign population in Italy at
the end of 2001. The histograms shown in the right panels of Figure
\ref{blockfigure} have been obtained by summing the draws from the
posterior distributions of $N_{l}$ for each block with at least 2
records in both the surveys. In fact, smaller blocks tend to produce
quite diffuse posterior distributions, making the MCMC
inference difficult without introducing a more concentrated prior. To overcome
this problem, the population size for the smaller blocks has been
fixed equal to $\hat{N}_{l}=(n^A_{l}+1)(n^B_{l}+1)/(\hat{T}_{l}+1)$
with $\hat{T}_{l}$ estimated by the classical approach. In particular,
one can notice that accounting for matching uncertainty with the
Jaro constrained model (central right panel) produces both a larger
estimate and larger uncertainty with respect to our approach (upper
right panel).\looseness=-1

\section{Simulation studies}
\label{sim}
We now evaluate our hierarchical model via a simulation study.
Artificial data are often used to evaluate record linkage techniques,
especially in computer science literature; see, for example,
\citet{Chr05} and \citet{Chr09}. Here, we consider three main different
simulation scenarios generating, at the superpopulation level, three
and six independent key variables (scenarios 1 and 3) and three
dependent key variables (scenario 2). Common features across different
simulations are as follows:
\begin{itemize}
\item the population size, fixed at $N=100$.
\item the sample size; we always assume $n^A=n^B$ equal to $70,80,90$.
\item the measurement error parameters $\beta$'s: their value has been
fixed at $(0.85,0.90,0.95)$.
\end{itemize}

In the first two scenarios, the three key variables assume,
respectively, 64, 16 and 4 categories, leading to a contingency table
with 4096 entries. In the independence case the means of the
population frequencies $F_j$ have been set equal to
$\theta_j=\theta_{j_1,j_2,j_3}=\prod_{i=1}^33 b_{j_i}$ where
$b_{j_i}\propto j_i$ with $j_i=1,\ldots, k_i$ for $i=1,2,3$. Under the
dependence model we set
$\theta_j=\theta_{j_1,j_2,j_3}=b_{j_3}b_{j_2|j_3} b_{j_1|j_3}$ where
$=b_{j_3}\propto j_3$, $b_{j_2|j_3}\propto j_2^{j_3}$ and
$b_{j_1|j_3}\propto j_1^{1/j_3}$. Finally, in the third scenario, the
6 key variables assume, respectively, 32, 16, 4, 4, 2 and 2 categories,
leading to a contingency table with 32,768 cells with
$\theta_j=\theta_{j_1,\ldots,j_6}=\prod_{i=1}^6 b_{j_i}$ where
$b_{j_i}\propto j_i$.

\begin{table}
\tabcolsep=0pt
\caption{Simulation study for evaluating the posterior mean $E(N)$
and the 95\% credible interval under the~new model ($M_1$), the Jaro
constrained model ($M_2$) and the Jaro unconstrained hybrid~approach~($M_3$)}\label{table3}
\begin{tabular*}{\tablewidth}{@{\extracolsep{4in minus 4in}}lcd{5.3}cd{5.3}ccccd{5.3}d{4.3}@{}}
\hline
&& \multicolumn{3}{c}{$\bolds{E(N)}$} &
\multicolumn{3}{c}{\textbf{Coverage}} &
\multicolumn{3}{c@{}}{\textbf{Length}} \\[-5pt]
&&\multicolumn{3}{c}{\hrulefill}&\multicolumn{3}{c}{\hrulefill}&\multicolumn{3}{c@{}}{\hrulefill}\\
$\bolds{\beta_i}$ & $\bolds{n^S}$ & \multicolumn{1}{c}{$\bolds{M_1} $} & $\bolds{M_2} $ & \multicolumn{1}{c}{$\bolds{M_3} $}&
$\bolds{M_1} $ & $\bolds{M_2} $ & $\bolds{M_3} $&
$\bolds{M_1} $ & \multicolumn{1}{c}{$\bolds{M_2} $} & \multicolumn{1}{c@{}}{$\bolds{M_3} $}
\\
\hline
\multicolumn{11}{@{}c@{}}{Scenario 1: independence with 3 key variables}\\
0.95
& 90 & 101\ (0.40) & 102\ (0.38) & 103\ (0.85) & 0.92 & 0.96 & 0.21 &
16\ (0.48) & 18\ (0.50) & 6\ (0.45) \\
& 80 & 99\ (0.55) & 103\ (0.68) & 98\ (0.93) & 0.96 & 0.96 & 0.30 &
24\ (0.65) & 28\ (0.81) & 9\ (0.53) \\
& 70 & 96\ (0.76) & 101\ (0.85) & 92\ (1.18) & 0.95 & 0.97 & 0.29 & 35\ (0.81) & 38\ (0.80) & 12\ (0.78) \\[3pt]
0.90 & 90 & 103\ (0.55) & 107\ (0.84) &103\ (1.10)& 0.97 & 0.91 & 0.20 &
26\ (0.83) & 32\ (1.30) & 6\ (0.60) \\
& 80 &100\ (0.78) &110\ (1.02) & 98\ (1.09) & 0.96 & 0.91 & 0.22 & 36\ (1.06) & 48\ (1.35) & 9\ (0.65) \\
& 70 & 96\ (1.06) &108\ (1.35) & 90\ (1.14) & 0.93 & 0.93 & 0.20 & 50\ (1.52) & 62\ (1.57) & 11\ (0.71)\\[3pt]
0.85 & 90 & 104\ (0.72) & 115\ (1.52) & 102\ (1.22) & 0.99 & 0.79 & 0.18 &37\ (1.32) & 56\ (2.39) &6\ (0.71)\\
& 80 & 100\ (0.87) & 116\ (1.53) & 93\ (0.94) & 0.99 & 0.93 & 0.17 &
51\ (1.49) & 75\ (2.61) & 6\ (0.46) \\
& 70 & 97\ (1.37) & 120\ (2.10) & 86\ (1.24) & 0.99 & 0.94 & 0.11 & 69\ (2.87) & 101\ (4.30) & 9\ (0.84) \\[6pt]
\multicolumn{11}{@{}c@{}}{Scenario 2: dependence with 3 key variables}\\
0.95 & 90 & 101\ (0.40) & 103\ (0.50) & 99\ (0.31) & 0.90 & 0.90 & 0.26 &
16\ (0.49) & 19\ (0.56) & 4\ (0.15)\\
& 80 & 99\ (0.59) & 103\ (0.61) & 93\ (0.52) & 0.96 & 0.96 & 0.24 &
26\ (0.58) & 29\ (0.67) & 6\ (0.29)\\
& 70 & 95\ (0.74) & 101\ (0.81) & 85\ (0.59) & 0.94 & 0.97 & 0.07 & 36\ (1.02) & 39\ (0.97) & 8\ (0.37) \\[3pt]
0.90 & 90& 102\ (0.49) & 111\ (0.74) & 99\ (0.45) & 0.96 & 0.82 & 0.23 &
27\ (0.74) & 37\ (1.05) & 4\ (0.22) \\
& 80& 100\ (0.74) & 111\ (1.07) & 92\ (0.84) & 0.94 & 0.90 & 0.15 &
38\ (1.09) & 50\ (1.49) & 6\ (0.49) \\
& 70& 93\ (0.96) & 107\ (1.26) & 85\ (1.16) & 0.92 & 0.93 & 0.08 & 45\ (1.41) & 60\ (1.79) & 8\ (0.84) \\[3pt]
0.85 & 90 & 104\ (0.66) & 120\ (1.27) & 99\ (0.75) & 0.98 & 0.74 & 0.21 &
38\ (1.21) & 58\ (2.16) & 4\ (0.40) \\
& 80 & 100\ (1.05) & 122\ (1.89) & 90\ (0.71) & 0.98 & 0.82 & 0.14 &
51\ (1.92) & 77\ (2.82) & 5\ (0.41) \\
& 70 & 95\ (1.21) & 123\ (1.84) & 82\ (0.74) & 0.98 & 0.94 & 0.05 & 63\ (2.30) & 104\ (3.69) & 6\ (0.48)\\[6pt]
\multicolumn{11}{@{}c@{}}{Scenario 3: independence with 6 key variables}\\
0.95 & 90& 101\ (0.27) & 101\ (0.29) & 102\ (0.44) & 0.83 & 0.83 & 0.27 &
10\ (0.25) & 11\ (0.28) & 5\ (0.23)\\
& 80 &101\ (0.45) & 102\ (0.51) & 100\ (0.64) & 0.93 & 0.95 & 0.59 &
19\ (0.51) & 20\ (0.47) & 10\ (0.37)\\
& 70 & 99\ (0.72) & 101\ (0.74) & 95\ (0.79) & 0.94 & 0.94 & 0.59 & 28\ (0.65) & 30\ (0.67) & 14\ (0.72)\\[3pt]
0.90 & 90 & 103\ (0.40) & 104\ (0.54) & 102\ (0.56) & 0.90 & 0.87 & 0.26 & 17\ (0.44) & 20\ (0.65) & 5\ (0.31) \\
& 80 & 100\ (0.66) & 104\ (0.83) & 95\ (0.63) & 0.98 & 0.93 & 0.39 &
26\ (0.80) &32\ (1.00) & 8\ (0.35)\\
& 70 & 98\ (0.94) & 104\ (0.11) &89\ (0.10) & 0.95 & 0.91 & 0.24 & 40\ (1.11) & 45\ (1.24) & 11\ (0.58)\\[3pt]
0.85 & 90& 105\ (0.65) & 111\ (0.99) &101\ (0.59) & 0.88 & 0.82 & 0.17 & 29\ (0.94)& 40\ (1.61) & 5\ (0.29)\\
& 80 & 100\ (0.95) & 112\ (1.65) & 93\ (0.85) & 0.93 & 0.85 & 0.15 &
38\ (1.34) & 54\ (2.42) & 6\ (0.47)\\
& 70 & 100\ (1.22) &118\ (1.85) & 85\ (0.91) & 0.98 & 0.89 & 0.14 & 58\ (1.96) & 81\ (3.10) & 8\ (0.56) \\
\hline
\end{tabular*}
\legend{Each cell reports a mean obtained with 100 pairs of data sets drawn
from our model with $N=100$, $n^A=n^B=n^S$ and $\beta_1=\cdots=\beta
_i=\cdots=\beta_h$. Standard errors are in parentheses.}
\end{table}

For each combination of model parameters, we have generated 100 pairs
of data sets. Each pair of data sets has been analyzed using our
hierarchical model with 45,000 iterations of the MCMC algorithm and
5000 iterations discarded for burn-in. For each pair of data sets we
also implemented the Jaro constrained model and the hybrid strategy
described in the previous sections. Mixing and convergence rates were
satisfactory based on the examination of trace plots.

In Table \ref{table3} we focus on the inference for $N$. For each of the three
approaches and for each group of 100 pairs of\vadjust{\goodbreak} data sets we report the
average values of the posterior mean of $N$, the estimated coverage of
the 95\% credibility intervals and their mean length. Estimated
standard errors are also given in parentheses. Note that in our
approach the average value of the posterior mean for $N$ is, in almost
every experimental condition, the closest to the true value $N=100$
and the one with the smallest standard error. However, the reduced
bias of our approach was to be expected because the simulation
generating process is exactly part of our model, while the other
approaches present several misspecification elements. Note also that
the Jaro constrained model and the hybrid approach have different
behaviours, the former overestimating $N$ and the latter
underestimating it. This is the same trend already observed in the
multiple block application.

The performance of the alternative approaches does not improve when
considering the interval estimates. In fact, with few exceptions, our
approach produces the interval estimates with a coverage level closest
to the nominal one. The hybrid approach, as expected, has dramatically
low coverage level since it does not account for matching
uncertainty. The Jaro constrained model always produces interval estimates
wider than those provided by our model, partly because it
only retrieves from the data the marginal information given by the
comparisons.

It is also interesting to note the behavior of the estimates with
respect to the information carried by the data. When the sample sizes
or the number of key variables increase, the uncertainty about $N$
reduces with all three methods. In addition, both our model and
the Jaro constrained model show a decrement in uncertainty as the
measurement error level decreases, that is, when the $\beta_i$'s
approach 1.

\begin{table}
\tabcolsep=0pt
\caption{Simulation study for evaluating the false match rates under
the new model (columns $M_1$), the Jaro constrained model (columns
$M_2$) and the Jaro unconstrained hybrid approach (columns $M_3$)}\label{table4}
\vspace*{-5pt}
\begin{tabular*}{\tablewidth}{@{\extracolsep{\fill}}lccccccc@{}}
\hline
&&\multicolumn{3}{c}{\textbf{FMR1}} &
\multicolumn{3}{c@{}}{\textbf{FMR2}} \\[-5pt]
&&\multicolumn{3}{c}{\hrulefill}&\multicolumn{3}{c@{}}{\hrulefill}\\
$\bolds{\beta_i}$ & $\bolds{n^S}$ & $\bolds{M_1} $ & $\bolds{M_2} $ & $\bolds{M_3} $&
$\bolds{M_1} $ & $\bolds{M_2} $ & $\bolds{M_3} $\\
\hline
\multicolumn{8}{@{}c@{}}{Scenario 1: independence with 3 key variables}\\
0.95 & 90 & 0.063 (0.004) & 0.052 (0.003) & 0.101 (0.004) & 0.129 (0.004) &
0.111 (0.005)& 0.126 (0.005) \\
& 80 & 0.074 (0.004) & 0.058 (0.004) & 0.146 (0.007) & 0.147
(0.006) & 0.148 (0.006) & 0.125 (0.006) \\
& 70 & 0.085 (0.005) & 0.073 (0.005) & 0.203 (0.008) & 0.165 (0.006) &
0.168 (0.006) & 0.130 (0.006) \\[3pt]
0.90 & 90& 0.095 (0.004) & 0.088 (0.004) & 0.185 (0.006) & 0.240 (0.007) &
0.244 (0.008) & 0.216 (0.006) \\
& 80& 0.100 (0.005) &0.090 (0.005) & 0.244 (0.007) & 0.274 (0.007)
& 0.286 (0.007)& 0.212 (0.006) \\
& 70& 0.123 (0.006) & 0.110 (0.006) & 0.319 (0.009) & 0.293 (0.007) &
0.309 (0.008) & 0.227 (0.007) \\[3pt]
0.85 & 90 & 0.130 (0.006) & 0.122 (0.006) & 0.307 (0.007) & 0.401 (0.007) &
0.401 (0.008) & 0.316 (0.007) \\
& 80 & 0.131 (0.005) & 0.122 (0.006) & 0.373 (0.008) & 0.423
(0.008) & 0.429 (0.007) &0.320 (0.009) \\
& 70 & 0.160 (0.007) & 0.144 (0.008) & 0.420 (0.009) & 0.447 (0.010) &
0.457 (0.010) & 0.322 (0.007) \\[6pt]
\multicolumn{8}{@{}c@{}}{Scenario 2: dependence with 3 key variables}\\
0.95& 90 & 0.065 (0.003) & 0.054 (0.003) & 0.138 (0.004) & 0.137 (0.005) &
0.126 (0.005) & 0.123 (0.005) \\
 & 80 & 0.075 (0.004) & 0.067 (0.005) & 0.205 (0.005) & 0.173
(0.006) & 0.152 (0.005) & 0.144 (0.006) \\
& 70 &0.083 (0.005) & 0.087 (0.005) & 0.278 (0.006) & 0.184 (0.006) &
0.170 (0.006) & 0.144 (0.006) \\[3pt]
0.90& 90 & 0.093 (0.004) & 0.091 (0.004) & 0.234 (0.007) & 0.270 (0.006) &
0.268 (0.006) & 0.227 (0.006) \\
 & 80 & 0.108 (0.005) & 0.108 (0.005) & 0.295 (0.008) & 0.289
(0.007) & 0.283 (0.008) & 0.228 (0.008) \\
& 70 & 0.117 (0.006) & 0.123 (0.007) & 0.353 (0.008) & 0.308 (0.007) &
0.298 (0.008) & 0.234 (0.006) \\[3pt]
0.85& 90 & 0.128 (0.005) & 0.145 (0.005) & 0.339 (0.007)& 0.426 (0.007) &
0.405 (0.007) &0.333 (0.007) \\
 & 80 & 0.141 (0.007) & 0.139 (0.007) & 0.398 (0.007) &
0.438 (0.007) & 0.422 (0.008) & 0.334 (0.007) \\
& 70 & 0.145 (0.007) & 0.149 (0.008) & 0.454 (0.007) & 0.463 (0.010) &
0.454 (0.010) & 0.331 (0.008) \\[6pt]
\multicolumn{8}{@{}c@{}}{Scenario 3: independence with 6 key variables}\\
0.95& 90 & 0.034 (0.002) & 0.030 (0.002) & 0.043 (0.003) & 0.054 (0.003) &
0.046 (0.003) & 0.058 (0.003) \\
 & 80 & 0.043 (0.003) & 0.040 (0.003) & 0.079 (0.005) & 0.065
(0.003)& 0.065 (0.003) & 0.070 (0.004) \\
& 70 & 0.068 (0.004) & 0.061 (0.004) & 0.134 (0.006) & 0.078 (0.005)
&0.083 (0.005)& 0.073 (0.004)\\[3pt]
0.90 & 90 & 0.071 (0.003) & 0.063 (0.003) & 0.115 (0.004) & 0.143 (0.005) &
0.140 (0.005) & 0.124 (0.005) \\
 & 80 & 0.089 (0.004) & 0.078 (0.003) & 0.175 (0.006) & 0.158
(0.005) & 0.168 (0.005) & 0.132 (0.005) \\
& 70& 0.104 (0.005) & 0.097 (0.005) & 0.244 (0.007) & 0.190 (0.006) &
0.204 (0.007) & 0.149 (0.006) \\[3pt]
0.85 & 90 & 0.126 (0.005) & 0.108 (0.004) & 0.232 (0.006) & 0.287 (0.006) &
0.297 (0.007) & 0.235 (0.006) \\
 & 80 & 0.142 (0.006) & 0.129 (0.006) & 0.297 (0.008) & 0.308
(0.009) & 0.330 (0.010) & 0.241 (0.008) \\
& 70 & 0.151 (0.007) & 0.131 (0.006) & 0.373 (0.008) & 0.342 (0.009) &
0.380 (0.011) & 0.256 (0.008) \\
\hline
\end{tabular*}
\legend{Each
cell reports a mean obtained with 100 pairs of data sets drawn
from our model with $N=100$, $n^A=n^B=n^S$ and $\beta_1=\cdots=\beta
_i=\cdots=\beta_h$. Standard errors are in parentheses.}
\end{table}

In Table \ref{table4} we report the results regarding the estimation of the
matching matrix~$C$. In particular, for each method we show the
average value of the False Match Rates defined by
\[
\mathrm{FMR1}=\frac{\sum_{ab} \hat{C}_{ab}(1-C_{ab})}{\sum_{ab} \hat{C}_{ab}}
\quad \mbox{and}\quad  \mathrm{FMR2}=\frac{\sum_{ab}
{C}_{ab}(1-\hat{C}_{ab})}{\sum_{ab} C_{ab}},
\]
where $\hat{C}$ is the
point estimate obtained using the quadratic loss. The results of the
comparisons among different methods would depend upon which type of
FMR is used. In particular, under the $\mathrm{FMR}1$ criterion, the better
performance is established by the Jaro constrained model, followed by
our approach and by the hybrid approach. However, one should recall
that the~Ja\-ro constrained model tends to overestimate $N$ and,
consequently, it leads to a~potential under-estimation of $T=\sum_{ab}
C_{ab}$. This way, the $\mathrm{FMR}1$ criterion would prefer the Jaro
constrained approach, because of its ``conservative behavior.'' From
our perspective this is another argument in favor of the inadequacy
of FMR1 as a single measure of performance of record linkage
procedures. Finally, note that, when using the $\mathrm{FMR}2$ criterion, the
hybrid approach quite often shows the better performance with our
model producing a lower rate than the Jaro constrained model under the
independence assumption.\vspace*{5pt}

\section{Discussion}
\label{disc}
Record linkage techniques pose several interesting problems both from
the methodological and the computational viewpoint. From a
methodological perspective, the definition itself of the statistical
framework within which comparisons among records should be performed is
still under debate: in this paper we have proposed a novel Bayesian
methodology.\looseness=1

While it is definitely true that the result of a statistical analysis
produced by an official organism must be objective (or --- at least --- it
should be perceived as such by the users), it is also undeniable [see
\citet{fienberg10}] that Bayesian ideas and techniques can play an
important role in official statistics, especially when important prior
(or extra-experimental) information about the variables of interest
exists and cannot be adequately exploited in a classical inference
framework. In addition, even when prior information is lacking, a
Bayesian analysis may be necessary simply because a classical approach
cannot provide answers without introducing strong assumptions, not
easily testable. In these situations a Bayesian analysis allows us, at
least, to perform a sensitivity analysis, with the aim of quantifying
the influence of the assumptions on inferences.

From a computational perspective record linkage problems be\-come
formi\-dable as soon as the sizes of the files are large. The
intensive simulation methods required by any Bayesian approach for a
matching problem make the computational problems in real applications
even more crucial. One of the most popular solutions, valid also for
our approach, is to perform the record linkage only between those
records which show the same values on some blocking variables which
are assumed to be recorded without errors. In addition, parallel
computations for separated blocks may reduce the computing time in a
significant way.

The proposed model is built up on the actually observed categorical
variables drawn from a finite population and no reduction of the
available information, for example, by using Boolean comparison
vectors, takes place. We also stress that prior information, provided
by experts or by previous surveys, can be introduced naturally into
the record linkage process via the superpopulation model, for
example, by giving specific association patterns between the key
variables. Another important benefit is the acknowledgment and
incorporation of the matching process uncertainty in estimating the
population size as well as other population parameters. At the same
time, the information available about the population parameters and
their uncertainty are accounted for in the record linkage.

Throughout the paper we have made some specific assumptions, such as
the fixed
sample sizes or the uniform distribution for the misspecified record
fields and their conditional independence given the true
values. Anyway, we are confident that our framework may provide a
basis for several extensions with more general assumptions. In
particular, some of the capture--recapture models used for a closed
population [see, e.g., \citet{wolter} and \citet{fien99} or
\citet{eros} and \citet{mvf08} for more advanced proposals] could be
incorporated as sampling models for the sample sizes and the number of
recaptures. Multiple recaptures could be handled following
\citet{ruffieuxgreen}, where a method for aligning multiple unlabeled
configurations has been proposed. By assuming an exchangeable prior for
$\beta_1,\ldots,\beta_h$, we may also remove the assumption of
conditional independence for the measurement error among record
fields. In addition, different measurement error probabilities across
files may be considered. Note also that the model has been developed
so that each block is separately evaluated. However, following
\citet{larsen05}, we could allow a ``borrowing of strength'' effect
across the blocks by introducing some extra layers in our prior
modeling. Some of these extensions will be the object of future
research. A similar approach for handling multivariate normal data is
discussed in \citet{litancwp09}.

An important aspect of record linkage procedures which we have not
addressed here is that of the nonrandomness of the samples, for
example, in applications using administrative lists provided by
register offices. This issue has some consequences in every modeling
approach to record linkage; however, discussion about these problems
is beyond the scope of this paper. In any case, we believe that the
idea of a Bayesian superpopulation model generating the lists might
be useful in this context too.

Finally, note that the computer science literature on record linkage
(also known as data matching or entity resolution) has developed, in
recent times, some impressive algorithms based on machine learning and
graph-based matching. Some relevant papers are \citet{Bha07} and
\citet{Kal06} and it would be interesting to compare these or similar
approaches with the statistical models presented in this
paper.

\begin{appendix}
\section{}\label{appendixA}
\vspace*{4pt}
The sampling models (\ref{mu.indipendenza}) and
(\ref{mu.marginale}) can be obtained as the marginal distribution of
$p(\mu^A,\mu^B,C,t|F)=p(\mu^A,\mu^B|C,t,F)p(C,t|F)$.
First, we average out $C$, so
$p(\mu^A,\mu^B|t,F)=\sum_{C} p(\mu^A,\mu^B,C|t,F)$.
In this sum we only need to consider those matrices $C$ with exactly
$t_j$ matches in the block
$\{(a,b): \mu^A_a=\mu^B_b=v_j\}$ for $j=1,\ldots,k$. The total number
of such \vspace*{2pt} matrices is $\prod_{j=1}^k \binom{f_j^A}{t_j} \binom
{f_j^B}{t_j} t_j!$. Also,
\begin{eqnarray*}
p(\mu^A,\mu^B|t,F)&=&\sum_{C} p(\mu^A,\mu^B,C|t,F)=\sum_{C} p(\mu
^A,\mu
^B|C,t,F)p(C|t,F)\\[2pt]
&=&p(\mu^A,\mu^B|C,t,F)p(C|t,F)\prod_{j=1}^k \pmatrix{f_j^A\cr t_j}
\pmatrix
{f_j^B\cr t_j} t_j!\\[2pt]
&=&\frac{\prod_{j=1}^k
\binom{F_j-t_j}{f_j^A-t_j,f_j^B-t_j,F_j-f_j^A-f_j^B+t_j}}
{\binom{N-T}{n^A-T,n^B-T,N-n^A-n^B+T}} \frac{\prod_{j=1}^k f_j^A f_j^B}{n^A!
n^B!}\\[2pt]
&=& \frac{1}{ \binom{n^A}{f_1^A,\ldots,f^A_k } }\frac{1}{\binom
{n^B}{f_1^B,\ldots,f^B_k}} \frac{ \prod_{j=1}^k
\binom{F_j-f_j^A}{f_j^B-t_j}
\binom{F_j-t_j}{f_j^A-t_j} }
{\binom{N-n^A}{n^B-T} \binom{N-T}{n^A-T}}.\vspace*{2pt}
\end{eqnarray*}
Then,
\begin{eqnarray*}
p(\mu^A,\mu^B|F)&=&\sum_t p(\mu^A,\mu^B,t|F)=\sum_t p(\mu^A,\mu
^B|t,F)p(t|F)\\[2pt]
&=& \frac{1}{ \binom{n^A}{f_1^A,\ldots,f^A_k } }\frac{1}{\binom
{n^B}{f_1^B,\ldots,f^B_k}} \sum_t\frac{ \prod_{j=1}^k
\binom{F_j-f_j^A}{f_j^B-t_j}
\binom{F_j-t_j}{f_j^A-t_j} }
{\binom{N-n^A}{n^B-T} \binom{N-T}{n^A-T}}
\frac{\binom{F_j}{t_j}}{\binom{N}{T}}
\frac{\binom{n^A}{T} \binom{N-n^A}{n^B-T}}{\binom{N}{n^B}}\\[2pt]
&=&\frac{1}{ \binom{n^A}{f_1^A,\ldots,f^A_k } }\frac{1}{\binom
{n^B}{f_1^B,\ldots,f^B_k}} \frac{1}{\binom{N}{n^A} \binom{N}{n^B}}\\[2pt]
&&{}\times\sum
_t \prod_{j=1}^k
\frac{F_j!}{(F_j-f_j^A-f_j^B+t_j)! (f_j^A-t_j)! (f_j^B-t_j)! t_j!}\\[2pt]
&=&
\frac{1}{ \binom{n^A}{f_1^A,\ldots,f^A_k } }\frac{1}{\binom
{n^B}{f_1^B,\ldots,f^B_k}} \frac{\prod_{j=1}^k
\binom{F_j}{f_j^A}
\binom{F_j}{f_j^B}}
{\binom{N}{n^A} \binom{N}{n^B}}.\vspace*{2pt}
\end{eqnarray*}

\section{}\label{appendixB}
The derivation of the full conditional distribution of
$t|F,\mu^A,\mu^B,\theta,\beta,x^A,x^B$:
\begin{eqnarray*}
&&p(t|F,\mu^A,\mu^B,\theta,\beta,x^A,x^B)\\ 
&&\qquad \propto p(\mu^A,\mu
^B|F,t)p(t|F)\\[-1.5pt]
\nonumber
&&\qquad \propto \frac{\prod_{j=1}^k \binom{F_j-t_j}{f_j^A-t_j}}{\binom
{N-T}{n^A-T}} \frac{\prod_{j=1}^k \binom
{F_j-f_j^A}{f_j^B-t_j}}{\binom
{N-n^A}{n^B-T}} \frac{\prod_{j=1}^k \binom{F_j}{t_j}}{\binom{N}{T}}
\frac{ \binom{n^A}{T} \binom{N-n^A}{n^B-T}}{\binom{N}{n^B}}\\[-1.5pt]
\nonumber
&&\qquad \propto \frac{\prod_{j=1}^k (\frac{F_j!}{t_j! (f_j^A-t_j)!
(f_j^B-t_j)! (F_j-f_j^A-f_j^B+t_j)} )}{\binom{N}{n^A}\binom{N}{n^B}}\\[-1.5pt]
&&\qquad \propto \prod_{j=1}^k \frac{
\binom{f_j^A}{t_j} \binom{F_j-f_j^A}{f_j^B-t_j}
}{\binom{F_j}{f_j^B}} .\vspace*{-1.5pt}
\end{eqnarray*}
\end{appendix}

\section*{Acknowledgments}
We would like to thank the referees, the Associate Editor and the
Editor (Professor Stephen E. Fienberg) for their very helpful comments
and suggestions that substantially improved a previous version of this
manuscript.

\begin{supplement}
\stitle{Data files and codes}\label{suppA}
\slink[doi]{10.1214/10-AOAS447SUPP}
\slink[url]{http://lib.stat.cmu.edu/aoas/447/supplement.zip}
\sdatatype{.zip}
\sdescription{Included in the supplementary\vadjust{\goodbreak}
material there are the following files: \mbox{exampleA.dat}, \mbox{exampleB.dat} and
exampleV.dat contain the data used in Section \ref{appl}. The files
B.Cat.matching.example.R, example.R, functions.r, gibbs.c contain the
codes. The file supplementary\_figure.pdf shows the trace plots for the
application described in Section \ref{appl}.}
\end{supplement}


%

\printaddresses


\begin{thebibliography}{61}

\bibitem[\protect\citeauthoryear{Albert and Dood}{2004}]{albertdood}
%
\begin{barticle}[author]
\bauthor{\bsnm{Albert},~\bfnm{P.~S.}\binits{P.~S.}} \AND
\bauthor{\bsnm{Dood},~\bfnm{L.~E.}\binits{L.~E.}}
(\byear{2004}).
\btitle{A cautionary note on robustness of latent class models for estimating
diagnostic error without a gold standard}.
\bjournal{Biometrics}
\bvolume{60}
\bpages{427--\break 435}.
\end{barticle}
\endbibitem

\bibitem[\protect\citeauthoryear{Alleva, Fortini and Tancredi}{2007}]{aft07}
\begin{binproceedings}[author]
\bauthor{\bsnm{Alleva},~\bfnm{Giorgio}\binits{G.}},
\bauthor{\bsnm{Fortini},~\bfnm{Marco}\binits{M.}} \AND
\bauthor{\bsnm{Tancredi},~\bfnm{Andrea}\binits{A.}}
(\byear{2007}).
\btitle{The control of non-sampling errors on linked data: An
application on
population census}.
In \bbooktitle{Proceedings of the 2007 Intermediate Conference. Risk and
Prediction. Venice}.
\end{binproceedings}
%
\endbibitem

\bibitem[\protect\citeauthoryear{Armstrong and Mayda}{1993}]{arm93}
%
\begin{barticle}[author]
\bauthor{\bsnm{Armstrong},~\bfnm{J.}\binits{J.}} \AND
\bauthor{\bsnm{Mayda},~\bfnm{J.~E.}\binits{J.~E.}}
(\byear{1993}).
\btitle{{M}odel-based estimation of record linkage error rates}.
\bjournal{Survey Methodology}
\bvolume{19}
\bpages{137--147}.
\end{barticle}
%
\endbibitem

\bibitem[\protect\citeauthoryear{Belin and Rubin}{1995}]{br95}
%
\begin{barticle}[author]
\bauthor{\bsnm{Belin},~\bfnm{T.~R.}\binits{T.~R.}} \AND
\bauthor{\bsnm{Rubin},~\bfnm{D.~B.}\binits{D.~B.}}
(\byear{1995}).
\btitle{A method for calibrating false---match rates in record linkage}.
\bjournal{J. Amer. Statist. Assoc.}
\bvolume{90}
\bpages{694--707}.
\end{barticle}
%
\endbibitem

\bibitem[\protect\citeauthoryear{Benjamini and Hochberg}{1995}]{benhoc95}
%
\begin{barticle}[author]
\bauthor{\bsnm{Benjamini},~\bfnm{Y.}\binits{Y.}} \AND
\bauthor{\bsnm{Hochberg},~\bfnm{Y.}\binits{Y.}}
(\byear{1995}).
\btitle{Controlling the false discovery rate: A practical and powerful approach
to multiple testing}.
\bjournal{J. Roy. Statist. Soc. Ser. B}
\bvolume{57}
\bpages{289--300}.
\end{barticle}
%
\endbibitem

\bibitem[\protect\citeauthoryear{Bhattacharya and Getoor}{2007}]{Bha07}
%
\begin{barticle}[author]
\bauthor{\bsnm{Bhattacharya},~\bfnm{Indrajit}\binits{I.}} \AND
\bauthor{\bsnm{Getoor},~\bfnm{Lise}\binits{L.}}
(\byear{2007}).
\btitle{Collective entity resolution in relational data}.
\bjournal{ACM Transactions on Knowledge Discovery from Data (TKDD)}
\bvolume{1}
\bpages{Article No. 5}.
\end{barticle}
%
\endbibitem

\bibitem[\protect\citeauthoryear{Bishop, Fienberg and Holland}{1975}]{bishop}
%
\begin{bbook}[author]
\bauthor{\bsnm{Bishop},~\bfnm{Y.~M.~M.}\binits{Y.~M.~M.}},
\bauthor{\bsnm{Fienberg},~\bfnm{S.~E.}\binits{S.~E.}} \AND
\bauthor{\bsnm{Holland},~\bfnm{P.~W.}\binits{P.~W.}}
(\byear{1975}).
\btitle{Discrete Multivariate Analysis: Theory and Parctice}.
\bpublisher{MIT Press}, \baddress{Cambridge}.
\end{bbook}
%
\endbibitem

\bibitem[\protect\citeauthoryear{Christen}{2005}]{Chr05}
%
\begin{binproceedings}[author]
\bauthor{\bsnm{Christen},~\bfnm{Peter}\binits{P.}}
(\byear{2005}).
\btitle{Probabilistic data generation for deduplication and data linkage}.
In \bbooktitle{IDEAL'05}.
\bseries{LNCS}
\bvolume{3578}
\bpages{109--116}.
\bpublisher{Springer},
\baddress{Berlin}.
\end{binproceedings}
%
\endbibitem

\bibitem[\protect\citeauthoryear{Christen and Pudjijono}{2009}]{Chr09}
%
\begin{binproceedings}[author]
\bauthor{\bsnm{Christen},~\bfnm{Peter}\binits{P.}} \AND
\bauthor{\bsnm{Pudjijono},~\bfnm{Agus}\binits{A.}}
(\byear{2009}).
\btitle{Accurate synthetic generation of realistic personal information}.
In \bbooktitle{Pacific-Asia Conference on Knowledge Discovery and Data Mining
(PAKDD'09)}.
\bseries{LNAI}
\bvolume{5476}
\bpages{507--514}.
\bpublisher{Springer},
\baddress{Berlin}.
\end{binproceedings}
%
\endbibitem

\bibitem[\protect\citeauthoryear{Copas and Hilton}{1990}]{copashilton}
%
\begin{barticle}[author]
\bauthor{\bsnm{Copas},~\bfnm{J.~B.}\binits{J.~B.}} \AND
\bauthor{\bsnm{Hilton},~\bfnm{F.~J.}\binits{F.~J.}}
(\byear{1990}).
\btitle{Record linkage: Statistical models for matching computer records}.
\bjournal{J. Roy. Statist. Soc. Ser. A}
\bvolume{153}
\bpages{287--320}.
\end{barticle}
%
\endbibitem

\bibitem[\protect\citeauthoryear{Darroch}{1958}]{darroch}
%
\begin{barticle}[author]
\bauthor{\bsnm{Darroch},~\bfnm{J.~N.}\binits{J.~N.}}
(\byear{1958}).
\btitle{The multiple capture census {I}. Estimation of a closed population}.
\bjournal{Biometrika}
\bvolume{45}
\bpages{343--358}.
\end{barticle}
%
\endbibitem

\bibitem[\protect\citeauthoryear{DeGroot and Goel}{1980}]{degrootgoel}
%
\begin{barticle}[author]
\bauthor{\bsnm{DeGroot},~\bfnm{M.~H.}\binits{M.~H.}} \AND
\bauthor{\bsnm{Goel},~\bfnm{P.~K.}\binits{P.~K.}}
(\byear{1980}).
\btitle{Estimation of the correlation coefficient from a broken random sample}.
\bjournal{Ann. Statist.}
\bvolume{8}
\bpages{264--278}.
\end{barticle}
%
\endbibitem

\bibitem[\protect\citeauthoryear{Do, Mueller and Tang}{2005}]{domu}
%
\begin{barticle}[author]
\bauthor{\bsnm{Do},~\bfnm{K.~A.}\binits{K.~A.}},
\bauthor{\bsnm{Mueller},~\bfnm{P.}\binits{P.}} \AND
\bauthor{\bsnm{Tang},~\bfnm{F.}\binits{F.}}
(\byear{2005}).
\btitle{A {B}ayesian mixture model for differential gene expression}.
\bjournal{J. Roy. Statist. Soc. Ser. C.}
\bvolume{54}
\bpages{627--644}.
\end{barticle}
%
\endbibitem

\bibitem[\protect\citeauthoryear{Ericson}{1969}]{ericson69}
%
\begin{barticle}[author]
\bauthor{\bsnm{Ericson},~\bfnm{W.~A.}\binits{W.~A.}}
(\byear{1969}).
\btitle{Subjective Bayesian models in sampling finite populations}.
\bjournal{J.~Roy. Statist. Soc. Ser. B}
\bvolume{31}
\bpages{195--224}.
\end{barticle}
%
\endbibitem

\bibitem[\protect\citeauthoryear{Erosheva, Fienberg and Joutard}{2007}]{eros}
%
\begin{barticle}[author]
\bauthor{\bsnm{Erosheva},~\bfnm{E.}\binits{E.}},
\bauthor{\bsnm{Fienberg},~\bfnm{S.}\binits{S.}} \AND
\bauthor{\bsnm{Joutard},~\bfnm{C.}\binits{C.}}
(\byear{2007}).
\btitle{Describing disability through individual level mixture models for
multivariate binary data}.
\bjournal{Ann. Appl. Statist.}
\bvolume{1}
\bpages{502--537}.
\end{barticle}
%
\endbibitem

\bibitem[\protect\citeauthoryear{Fellegi and Sunter}{1969}]{fesu69}
%
\begin{barticle}[author]
\bauthor{\bsnm{Fellegi},~\bfnm{I.~P.}\binits{I.~P.}} \AND
\bauthor{\bsnm{Sunter},~\bfnm{A.~B.}\binits{A.~B.}}
(\byear{1969}).
\btitle{{A} theory of record linkage}.
\bjournal{J. Amer. Statist. Assoc.}
\bvolume{64}
\bpages{1183--1210}.
\end{barticle}
%
\endbibitem

\bibitem[\protect\citeauthoryear{Fienberg}{2011}]{fienberg10}
%
\begin{bmisc}[author]
\bauthor{\bsnm{Fienberg},~\bfnm{S.~E.}\binits{S.~E.}}
(\byear{2011}).
\bhowpublished{Bayesian models and methods in public policy and government settings.
\textit{Statist. Sci.} To appear.}
\end{bmisc}
%
\endbibitem

\bibitem[\protect\citeauthoryear{Fienberg, Johnson and Junker}{1999}]{fien99}
%
\begin{barticle}[author]
\bauthor{\bsnm{Fienberg},~\bfnm{Stephen~E.}\binits{S.~E.}},
\bauthor{\bsnm{Johnson},~\bfnm{M.~S.}\binits{M.~S.}} \AND
\bauthor{\bsnm{Junker},~\bfnm{B.~W.}\binits{B.~W.}}
(\byear{1999}).
\btitle{Classical multilevel and Bayesian approaches to population size
estimation using multiple list, part 3.}
\bjournal{J. Roy. Statist. Soc. Ser. A}
\bvolume{162}
\bpages{383--405}.
\end{barticle}
%
\endbibitem

\bibitem[\protect\citeauthoryear{Fienberg and Manrique-Vallier}{2009}]{mvf09}
%
\begin{barticle}[author]
\bauthor{\bsnm{Fienberg},~\bfnm{Stephen~E.}\binits{S.~E.}} \AND
\bauthor{\bsnm{Manrique-Vallier},~\bfnm{Daniel}\binits{D.}}
(\byear{2009}).
\btitle{Integrated methodology for multiple systems estimation and record
linkage using a missing data formulation}.
\bjournal{Advances in Statistical Analysis}
\bvolume{93}
\bpages{49--60}.
\end{barticle}\vadjust{\goodbreak}
%
\endbibitem

\bibitem[\protect\citeauthoryear{Forster and Webb}{2007}]{forsterwebb}
%
\begin{barticle}[author]
\bauthor{\bsnm{Forster},~\bfnm{J.~J.}\binits{J.~J.}} \AND
\bauthor{\bsnm{Webb},~\bfnm{E.~L.}\binits{E.~L.}}
(\byear{2007}).
\btitle{Bayesian disclosure risk assesment: Predicting small
frequencies in
contingency table}.
\bjournal{J. Roy. Statist. Soc. Ser. C}
\bvolume{56}
\bpages{551--570}.
\end{barticle}
%
\endbibitem

\bibitem[\protect\citeauthoryear{Fortini et~al.}{2001}]{Fortini}
%
\begin{barticle}[author]
\bauthor{\bsnm{Fortini},~\bfnm{M.}\binits{M.}},
\bauthor{\bsnm{Liseo},~\bfnm{B.}\binits{B.}},
\bauthor{\bsnm{Nuccitelli},~\bfnm{A.}\binits{A.}} \AND
\bauthor{\bsnm{Scanu},~\bfnm{M.}\binits{M.}}
(\byear{2001}).
\btitle{On {B}ayesian record linkage}.
\bjournal{Research in Official Statistics}
\bvolume{4}
\bpages{185--198}.
\end{barticle}
%
\endbibitem

\bibitem[\protect\citeauthoryear{Fortini et~al.}{2002}]{Fortini2}
%
\begin{binproceedings}[author]
\bauthor{\bsnm{Fortini},~\bfnm{M.}\binits{M.}},
\bauthor{\bsnm{Liseo},~\bfnm{B.}\binits{B.}},
\bauthor{\bsnm{Nuccitelli},~\bfnm{A.}\binits{A.}} \AND
\bauthor{\bsnm{Scanu},~\bfnm{M.}\binits{M.}}
(\byear{2002}).
\btitle{Modelling issues in record linkage: A {B}ayesian perspective}.
In \bbooktitle{Proceedings of the Section on Survey Research Methods}
\bpages{1008--1013}.
\bpublisher{Amer. Statist. Assoc.},
\baddress{Alexandria, VA}.
\end{binproceedings}
%
\endbibitem

\bibitem[\protect\citeauthoryear{Genovese and Wasserman}{2003}]{genowa01}
%
\begin{bincollection}[author]
\bauthor{\bsnm{Genovese},~\bfnm{C.}\binits{C.}} \AND
\bauthor{\bsnm{Wasserman},~\bfnm{L.}\binits{L.}}
(\byear{2003}).
\btitle{Bayesian and frequentist multiple testing (with discussion)}.
In \bbooktitle{Bayesian Statistics, 7 ({T}enerife, 2002)}
\bpages{145--161}.
\bpublisher{Oxford Univ. Press}, \baddress{New York}.
\bid{mr={2003171}}
\end{bincollection}
%
\endbibitem

\bibitem[\protect\citeauthoryear{Green and Mardia}{2006}]{greenmardia}
%
\begin{barticle}[author]
\bauthor{\bsnm{Green},~\bfnm{P.~J.}\binits{P.~J.}} \AND
\bauthor{\bsnm{Mardia},~\bfnm{K.~V.}\binits{K.~V.}}
(\byear{2006}).
\btitle{Bayesian alignment using hierarchical models, with applicationin
protein bioinformatics}.
\bjournal{Biometrika}
\bvolume{93}
\bpages{235--254}.
\end{barticle}
%
\endbibitem

\bibitem[\protect\citeauthoryear{Herzog, Scheuren and
Winkler}{2007}]{herzogetal}
%
\begin{bbook}[author]
\bauthor{\bsnm{Herzog},~\bfnm{T.~N.}\binits{T.~N.}},
\bauthor{\bsnm{Scheuren},~\bfnm{F.~J.}\binits{F.~J.}} \AND
\bauthor{\bsnm{Winkler},~\bfnm{W.~E.}\binits{W.~E.}}
(\byear{2007}).
\btitle{Data Quality and Record Linkage Techniques}.
\bpublisher{Springer},
\baddress{New York}.
\end{bbook}
%
\endbibitem

\bibitem[\protect\citeauthoryear{Hoadley}{1969}]{hoadly}
%
\begin{barticle}[author]
\bauthor{\bsnm{Hoadley},~\bfnm{B.}\binits{B.}}
(\byear{1969}).
\btitle{The compound multinomial distribution and Bayesian analysis of
categorical data from finite populations}.
\bjournal{J. Amer. Statist. Assoc.}
\bvolume{64}
\bpages{216--229}.
\end{barticle}
%
\endbibitem

\bibitem[\protect\citeauthoryear{Jaro}{1989}]{jaro89}
%
\begin{barticle}[author]
\bauthor{\bsnm{Jaro},~\bfnm{M.}\binits{M.}}
(\byear{1989}).
\btitle{{A}dvances in record-linkage methodology as applied to
matching the
1985 census of {T}ampa, {F}lorida}.
\bjournal{J. Amer. Statist. Assoc.}
\bvolume{84}
\bpages{414--420}.
\end{barticle}
%
\endbibitem

\bibitem[\protect\citeauthoryear{Judson}{2007}]{judson}
%
\begin{barticle}[author]
\bauthor{\bsnm{Judson},~\bfnm{D.~H.}\binits{D.~H.}}
(\byear{2007}).
\btitle{Information integration for constructing social statistics: History,
theory and ideas towards a reserach program}.
\bjournal{J. Roy. Statist. Soc. Ser. A}
\bvolume{170}
\bpages{483--501}.
\end{barticle}
%
\endbibitem

\bibitem[\protect\citeauthoryear{Kalashnikov and Mehrotra}{2006}]{Kal06}
%
\begin{barticle}[author]
\bauthor{\bsnm{Kalashnikov},~\bfnm{D.~V.}\binits{D.~V.}} \AND
\bauthor{\bsnm{Mehrotra},~\bfnm{S.}\binits{S.}}
(\byear{2006}).
\btitle{Domain-independent data cleaning via analysis of entity-relationship
graph}.
\bjournal{ACM Transactions on Database Systems}
\bvolume{31}
\bpages{716--767}.
\end{barticle}
%
\endbibitem

\bibitem[\protect\citeauthoryear{Kelley}{1986}]{Kelley}
%
\begin{binproceedings}[author]
\bauthor{\bsnm{Kelley},~\bfnm{P.}\binits{P.}}
(\byear{1986}).
\btitle{Robustness of the Census Bureau's record linkage system}.
In \bbooktitle{Proceedings of the Section on Survey Research Methods, American
Statistical Association}
\bpages{620--624}.
\end{binproceedings}
%
\endbibitem

\bibitem[\protect\citeauthoryear{Lahiri and Larsen}{2005}]{larsenlahiri}
%
\begin{barticle}[author]
\bauthor{\bsnm{Lahiri},~\bfnm{P.}\binits{P.}} \AND
\bauthor{\bsnm{Larsen},~\bfnm{M.~D.}\binits{M.~D.}}
(\byear{2005}).
\btitle{Regression analysis with linked data}.
\bjournal{J. Amer. Statist. Assoc.}
\bvolume{100}
\bpages{222--230}.
\end{barticle}
%
\endbibitem

\bibitem[\protect\citeauthoryear{Larsen}{1999}]{lar99}
%
\begin{binproceedings}[author]
\bauthor{\bsnm{Larsen},~\bfnm{M.~D.}\binits{M.~D.}}
(\byear{1999}).
\btitle{{M}ultiple imputation analysis of records linked using mixture model}.
In \bbooktitle{Proceedings of Survey Methods Section, Statistical
Society of
Canada}
\bpages{65--71}.
\bpublisher{Statistical Society of Canada}, \baddress{Ottawa}.
\end{binproceedings}
%
\endbibitem

\bibitem[\protect\citeauthoryear{Larsen}{2004}]{lar05}
%
\begin{bincollection}[author]
\bauthor{\bsnm{Larsen},~\bfnm{M.~D.}\binits{M.~D.}}
(\byear{2004}).
\btitle{Record {Linkage} using finite mixture models}.
In \bbooktitle{Applied {B}ayesian {M}odeling and {C}ausal {I}nference from
{I}ncomplete {D}ata {Perspectives}}
(\beditor{A. Gelman and X.-L. Meng}, eds.)
\bpages{309--318}.
\bpublisher{Wiley}, \baddress{New York}.
\bid{mr={2003171}}
\end{bincollection}
%
\endbibitem

\bibitem[\protect\citeauthoryear{Larsen}{2005}]{larsen05}
%
\begin{binproceedings}[author]
\bauthor{\bsnm{Larsen},~\bfnm{M.~D.}\binits{M.~D.}}
(\byear{2005}).
\btitle{Advances in record linkage theory: Hierarchical Bayesian
record linkage
theory}.
In \bbooktitle{Proceedings of the Section on Survey Research Methods}
\bpages{3277--3283}.
\bpublisher{Amer. Statist. Assoc.},
\baddress{Alexandria, VA}.
\end{binproceedings}
%
\endbibitem

\bibitem[\protect\citeauthoryear{Larsen and Rubin}{2001}]{laru01}
%
\begin{barticle}[author]
\bauthor{\bsnm{Larsen},~\bfnm{M.~D.}\binits{M.~D.}} \AND
\bauthor{\bsnm{Rubin},~\bfnm{D.~B.}\binits{D.~B.}}
(\byear{2001}).
\btitle{{I}terative automated record linkage using mixture models}.
\bjournal{J. Amer. Statist. Assoc.}
\bvolume{96}
\bpages{32--41}.
\end{barticle}
%
\endbibitem

\bibitem[\protect\citeauthoryear{Lindley}{1977}]{lindley77}
%
\begin{barticle}[author]
\bauthor{\bsnm{Lindley},~\bfnm{D.~V.}\binits{D.~V.}}
(\byear{1977}).
\btitle{A problem in forensic science}.
\bjournal{Biometrika}
\bvolume{64}
\bpages{207--213}.
\end{barticle}
%
\endbibitem

\bibitem[\protect\citeauthoryear{Link et~al.}{2009}]{link}
%
\begin{barticle}[author]
\bauthor{\bsnm{Link},~\bfnm{W.~A.}\binits{W.~A.}},
\bauthor{\bsnm{Yoshizaki},~\bfnm{J.}\binits{J.}},
\bauthor{\bsnm{Bailey},~\bfnm{L.~L.}\binits{L.~L.}} \AND
\bauthor{\bsnm{Pollock},~\bfnm{K.~H.}\binits{K.~H.}}
(\byear{2009}).
\btitle{Uncovering a~latent multinomial: {A}nalysis of {m}ark--{r}ecapture
{d}ata with {m}isidentication}.
\bjournal{Biometrics}
\bvolume{66}
\bpages{178--185}.
\end{barticle}
%
\endbibitem

\bibitem[\protect\citeauthoryear{Liseo and Tancredi}{2009}]{litancwp09}
%
\begin{bmisc}[author]
\bauthor{\bsnm{Liseo},~\bfnm{B.}\binits{B.}} \AND
\bauthor{\bsnm{Tancredi},~\bfnm{A.}\binits{A.}}
(\byear{2009}).
\btitle{Bayesian estimation of population size via linkage of multivariate
Normal data sets}.
\bhowpublished{Working Paper 66. Dept. Methods and Models for Economics,
Territory and Finance, Sapienza Univ. Rome}.
\end{bmisc}
%
\endbibitem

\bibitem[\protect\citeauthoryear{Manrique-Vallier and Fienberg}{2008}]{mvf08}
%
\begin{barticle}[author]
\bauthor{\bsnm{Manrique-Vallier},~\bfnm{Daniel}\binits{D.}} \AND
\bauthor{\bsnm{Fienberg},~\bfnm{Stephen~E.}\binits{S.~E.}}
(\byear{2008}).
\btitle{Population size estimation using individual level mixture models}.
\bjournal{Biom. J.}
\bvolume{50}
\bpages{1051--1063}.
\end{barticle}
%
\endbibitem

\bibitem[\protect\citeauthoryear{Marin and Robert}{2007}]{marinrobert}
%
\begin{bbook}[author]
\bauthor{\bsnm{Marin},~\bfnm{J.~M.}\binits{J.~M.}} \AND
\bauthor{\bsnm{Robert},~\bfnm{C.~P.}\binits{C.~P.}}
(\byear{2007}).
\btitle{Bayesian Core. A Practical Approach to Computational Statistics}.
\bpublisher{Springer},
\baddress{New York}.
\end{bbook}
%
\endbibitem

\bibitem[\protect\citeauthoryear{McGlincy}{2004}]{mg04}
%
\begin{binproceedings}[author]
\bauthor{\bsnm{McGlincy},~\bfnm{M.}\binits{M.}}
(\byear{2004}).
\btitle{A {B}ayesian record linkage methodology for multiple
imputation of
missing data}.
In \bbooktitle{Proceedings of the Section on Survey Research Methods}
\bpages{4001--4008}.
\bpublisher{Amer. Statist. Assoc.},
\baddress{Alexandria, VA}.
\end{binproceedings}
%
\endbibitem

\bibitem[\protect\citeauthoryear{Newcombe}{1967}]{newcombe67}
%
\begin{barticle}[author]
\bauthor{\bsnm{Newcombe},~\bfnm{H.~B.}\binits{H.~B.}}
(\byear{1967}).
\btitle{Record linkage: The design of efficient systems for linking records
into individual and family histories}.
\bjournal{American Journal of Human Genetics}
\bvolume{9}
\bpages{335--359}.
\end{barticle}
%
\endbibitem

\bibitem[\protect\citeauthoryear{Newcombe et~al.}{1959}]{newcombe59}
%
\begin{barticle}[author]
\bauthor{\bsnm{Newcombe},~\bfnm{H.~B.}\binits{H.~B.}},
\bauthor{\bsnm{Kennedy},~\bfnm{J.~M.}\binits{J.~M.}},
\bauthor{\bsnm{Axford},~\bfnm{S.~J.}\binits{S.~J.}} \AND
\bauthor{\bsnm{James},~\bfnm{A.~P.}\binits{A.~P.}}
(\byear{1959}).
\btitle{Automatic linkage of vital records}.
\bjournal{Science}
\bvolume{130}
\bpages{954--959}.
\end{barticle}
%
\endbibitem

\bibitem[\protect\citeauthoryear{Nor{\'{e}}n, Orre and Bate}{2005}]{noren05}
%
\begin{binproceedings}[author]
\bauthor{\bsnm{Nor{\'{e}}n},~\bfnm{G.~Niklas}\binits{G.~N.}},
\bauthor{\bsnm{Orre},~\bfnm{Roland}\binits{R.}} \AND
\bauthor{\bsnm{Bate},~\bfnm{Andrew}\binits{A.}}
(\byear{2005}).
\btitle{A hit--miss model for duplicate detection in the WHO drug safety
database}.
In \bbooktitle{KDD'05: Proceedings of the Eleventh ACM SIGKDD International
Conference on Knowledge Discovery in Data Mining}
\bpages{459--468}.
\baddress{Canada}.
\end{binproceedings}
%
\endbibitem

\bibitem[\protect\citeauthoryear{O'Hagan and Forster}{2004}]{ohaganforster}
%
\begin{bbook}[author]
\bauthor{\bsnm{O'Hagan},~\bfnm{A.}\binits{A.}} \AND
\bauthor{\bsnm{Forster},~\bfnm{J.}\binits{J.}}
(\byear{2004}).
\btitle{Kendall's Advanced Theory of Statistics. Volume 2B. Bayesian
Inference}.
\bpublisher{Arnold},
\baddress{London}.
\end{bbook}
%
\endbibitem

\bibitem[\protect\citeauthoryear{Pepe}{2003}]{pepe2003}
%
\begin{bbook}[author]
\bauthor{\bsnm{Pepe},~\bfnm{M.~S.}\binits{M.~S.}}
(\byear{2003}).
\btitle{The {S}tatistical {E}valuation of {M}edical {T}est for {C}lassification
and {P}rediction}.
\bpublisher{Oxford Univ. Press}, \baddress{London}.
\end{bbook}
%
\endbibitem

\bibitem[\protect\citeauthoryear{Pepe and Janes}{2007}]{pepejanes}
%
\begin{barticle}[author]
\bauthor{\bsnm{Pepe},~\bfnm{M.~S.}\binits{M.~S.}} \AND
\bauthor{\bsnm{Janes},~\bfnm{H.}\binits{H.}}
(\byear{2007}).
\btitle{Insights into latent class analysis of diagnostic test performance}.
\bjournal{Biostatistics}
\bvolume{8}
\bpages{474--484}.
\end{barticle}
%
\endbibitem

\bibitem[\protect\citeauthoryear{Perez et~al.}{2007}]{peirezetal}
%
\begin{barticle}[author]
\bauthor{\bsnm{Perez},~\bfnm{C.~J.}\binits{C.~J.}},
\bauthor{\bsnm{Giron},~\bfnm{F.~J.}\binits{F.~J.}},
\bauthor{\bsnm{Martin},~\bfnm{J.}\binits{J.}},
\bauthor{\bsnm{Ruiz},~\bfnm{M.}\binits{M.}} \AND
\bauthor{\bsnm{Rojano},~\bfnm{C.}\binits{C.}}
(\byear{2007}).
\btitle{Misclassified multinomial data: A {B}ayesian approach}.
\bjournal{Rev. R. Acad. Cien. Serie A. Mat}
\bvolume{101}
\bpages{71--80}.
\end{barticle}
%
\endbibitem

\bibitem[\protect\citeauthoryear{R Development Core Team}{2009}]{RRR}
%
\begin{bmisc}[author]
\borganization{R Development Core Team}
(\byear{2009}).
\bhowpublished{\textit{R: A Language and Environment for Statistical Computing}.
R Foundation for Statistical Computing, Vienna,
Austria}.
\end{bmisc}
%
\endbibitem

\bibitem[\protect\citeauthoryear{Robert and Casella}{2004}]{robertcasella04}
%
\begin{bbook}[author]
\bauthor{\bsnm{Robert},~\bfnm{C.~P.}\binits{C.~P.}} \AND
\bauthor{\bsnm{Casella},~\bfnm{G.}\binits{G.}}
(\byear{2004}).
\btitle{Monte Carlo Statistical Methods},
\bedition{2nd} ed.
\bpublisher{Springer},
\baddress{New York}.
\end{bbook}
%
\endbibitem

\bibitem[\protect\citeauthoryear{Ruffieux and Green}{2009}]{ruffieuxgreen}
%
\begin{barticle}[author]
\bauthor{\bsnm{Ruffieux},~\bfnm{Yann}\binits{Y.}} \AND
\bauthor{\bsnm{Green},~\bfnm{Peter~J.}\binits{P.~J.}}
(\byear{2009}).
\btitle{Alignment of multiple configurations using hierachical models}.
\bjournal{J. Comput. Graph. Statist.}
\bvolume{18}
\bpages{756--773}.
\end{barticle}
%
\endbibitem

\bibitem[\protect\citeauthoryear{Seber}{1986}]{seber86}
%
\begin{barticle}[author]
\bauthor{\bsnm{Seber},~\bfnm{G.~A.~F.}\binits{G.~A.~F.}}
(\byear{1986}).
\btitle{A review of estimating animal abundance}.
\bjournal{Biometrics}
\bvolume{42}
\bpages{267--292}.
\end{barticle}
%
\endbibitem

\bibitem[\protect\citeauthoryear{Swartz et~al.}{2004}]{swartzetal}
%
\begin{barticle}[author]
\bauthor{\bsnm{Swartz},~\bfnm{T.}\binits{T.}},
\bauthor{\bsnm{Haitovsky},~\bfnm{Y.}\binits{Y.}},
\bauthor{\bsnm{Vexler},~\bfnm{A.}\binits{A.}} \AND
\bauthor{\bsnm{Yang},~\bfnm{T.}\binits{T.}}
(\byear{2004}).
\btitle{Bayesian identifiability and misclassification in multinomial data}.
\bjournal{Canad. J. Statist.}
\bvolume{32}
\bpages{285--302}.
\end{barticle}
%
\endbibitem

\bibitem[\protect\citeauthoryear{Tancredi and Liseo}{2011}]{sptl}
%
\begin{bmisc}[author]
\bauthor{\bsnm{Tancredi},~\bfnm{Andrea}\binits{A.}} \AND
\bauthor{\bsnm{Liseo},~\bfnm{Brunero}\binits{B.}}
(\byear{2011}).
\bhowpublished{Supplement to ``A hierarchical Bayesian approach to record
linkage and
population size problems.'' DOI: \href{http://dx.doi.org/10.1214/10-AOAS447SUPP}{10.1214/10-AOAS447SUPP}.}
\end{bmisc}
%
\endbibitem

\bibitem[\protect\citeauthoryear{Winkler}{1993}]{wink93}
%
\begin{binproceedings}[author]
\bauthor{\bsnm{Winkler},~\bfnm{W.~E.}\binits{W.~E.}}
(\byear{1993}).
\btitle{{I}mproved decision rules in the {F}ellegi--{S}unter model of record
linkage}.
In \bbooktitle{Proceedings of the Section on Survey Research Methods}
\bpages{274--279}.
\bpublisher{Amer. Statist. Assoc.},
\baddress{Alexandria, VA}.
\end{binproceedings}
%
\endbibitem

\bibitem[\protect\citeauthoryear{Winkler}{1995}]{winkler95}
%
\begin{bincollection}[author]
\bauthor{\bsnm{Winkler},~\bfnm{W.~E.}\binits{W.~E.}}
(\byear{1995}).
\btitle{Matching and record linkage}.
In \bbooktitle{Buisness Survey Methods}
(\beditor{B.~G.~Cox, D.~A. Binder, B. N. Chinnappa, A. Christianson, M. J. Colledge
and P. S. Kott}, eds.)
\bpages{355--384}.
\bpublisher{Wiley}, \baddress{New York}.
\end{bincollection}
%
\endbibitem

\bibitem[\protect\citeauthoryear{Winkler}{2000}]{win00}
%
\begin{binproceedings}[author]
\bauthor{\bsnm{Winkler},~\bfnm{W.~E.}\binits{W.~E.}}
(\byear{2000}).
\btitle{Machine learning, information retrieval and record linkage}.
In \bbooktitle{Proceedings of the Section on Survey Research Methods}
\bpages{20--29}.
\bpublisher{Amer. Statist. Assoc.},
\baddress{Alexandria, VA}.
\end{binproceedings}
%
\endbibitem

\bibitem[\protect\citeauthoryear{Winkler}{2004}]{winkler04}
%
\begin{binproceedings}[author]
\bauthor{\bsnm{Winkler},~\bfnm{W.~E.}\binits{W.~E.}}
(\byear{2004}).
\btitle{Approximate string comparator search strategies for very large
administrative lists}.
In \bbooktitle{Proceedings of the Section on Survey Research Methods}
\bpages{4595--4602}.
\bpublisher{Amer. Statist. Assoc.},
\baddress{Alexandria, VA}.
\end{binproceedings}
%
\endbibitem

\bibitem[\protect\citeauthoryear{Wolter}{1986}]{wolter}
%
\begin{barticle}[author]
\bauthor{\bsnm{Wolter},~\bfnm{Kirk~M.}\binits{K.~M.}}
(\byear{1986}).
\btitle{Some coverage error models for census data}.
\bjournal{J. Amer. Statist. Assoc.}
\bvolume{81}
\bpages{338--345}.
\end{barticle}
%
\endbibitem

\bibitem[\protect\citeauthoryear{Wright et~al.}{2009}]{wright09}
%
\begin{barticle}[author]
\bauthor{\bsnm{Wright},~\bfnm{Janine~A.}\binits{J.~A.}},
\bauthor{\bsnm{Baker},~\bfnm{Richard~J.}\binits{R.~J.}},
\bauthor{\bsnm{Schofield},~\bfnm{Matthew~R.}\binits{M.~R.}},
\bauthor{\bsnm{Frantz},~\bfnm{Alain~C.}\binits{A.~C.}},
\bauthor{\bsnm{Byrom},~\bfnm{Andrea~E.}\binits{A.~E.}} \AND
\bauthor{\bsnm{Gleeson},~\bfnm{Dianne~M.}\binits{D.~M.}}
(\byear{2009}).
\btitle{Incorporating genotype uncertainty into mark-recapture-type
models for
estimating abundance using DNA samples}.
\bjournal{Biometrics}
\bvolume{65}
\bpages{833--840}.
\end{barticle}
%
\endbibitem

\end{thebibliography}
\end{document}